\newcommand{\sratio} {$\avg{S_1}/\avg{S_2}$}
\newcommand{\sratioe} {\avg{S_1}/\avg{S_2}}
\newcommand{\bara}{16~$\times$~16~$\times$~200~mm$^3$}
\newcommand{\barb}{20~$\times$~20~$\times$~200~mm$^3$}
\newcommand{\barc}{40~$\times$~40~$\times$~400~mm$^3$}
\newcommand{\avg}[1]{\left< #1 \right>} 
\newcommand{\sigARM}{$\sigma_{\mbox{\scriptsize ARM}}$}

\newcommand{\Cs}{$^{137}$Cs}
\newcommand{\Na}{$^{22}$Na}
\newcommand{\Sn}{$^{113}$Sn}

\newcommand{\gae}{\lower 2pt \hbox{$\, \buildrel {\scriptstyle >}\over {\scriptstyle
\sim}\,$}}
\newcommand{\lae}{\lower 2pt \hbox{$\, \buildrel {\scriptstyle <}\over {\scriptstyle
\sim}\,$}}

\newcommand{\Nbar}{${N_{\mbox{\scriptsize BAR}}}$}

\documentclass[final,3p,times]{elsarticle} 

\usepackage{placeins}

\usepackage{amssymb}

\usepackage[nodots,nocompress]{numcompress}

\usepackage{color}

\biboptions{square}

\journal{Preprint submitted to Nuclear Instruments and Methods in Physics Research A}

\begin{document}

\begin{frontmatter}


\title{Development of a Compton Imager Based on Bars of Scintillator}


\author[label1]{A.M.L. MacLeod\corref{cor1}\fnref{label4}}
\address[label1]{Physics Department, McGill University, 3600 rue University, Montreal, Quebec, Canada, H3A 2T8}
\ead{audrey.macleod@nrcan-rncan.gc.ca}
\cortext[cor1]{now at the Geological Survey of Canada.}
\author[label1]{P.J. Boyle}
\author[label1]{D.S. Hanna}
\author[label2]{P.R.B. Saull}
\author[label3]{L.E. Sinclair}
\author[label3]{H.C.J. Seywerd}
\address[label2]{Measurement Science and Standards, National Research Council, 1200 Montreal Road, Ottawa, Ontario, Canada, K1A OR6}
\address[label3]{Geological Survey of Canada, Natural Resources Canada, 601 Booth Street, Ottawa, Ontario, Canada, K1A 0E8}

\begin{abstract} 
We have developed a compact Compton gamma-ray imager with a large field of view and a low channel-count that is capable of quickly localizing gamma-ray sources in the few hundred keV -- several MeV range. The two detector planes (scatter and absorber) employ bars of NaI(Tl) read out by photomultiplier tubes (PMTs) located at each end. The long-range imaging performance has been tested from 392~keV to 1274~keV. An angular resolution measure of $2.72^{\circ}~\pm~0.06^{\circ}$ and an efficiency of $(1.79~\pm~0.04)~\times~10^{-3}$ at 662~keV is obtained. A \Cs~(662~keV) source equivalent to a 10~mCi source 40~m away can be located in 60~seconds with an uncertainty of about a degree. No significant degradation in imaging performance is observed for source angles up to 40$^{\circ}$ off axis. 
\end{abstract}

\begin{keyword}
Compton Imaging \sep
Compton Telescope \sep
NaI \sep
Scintillator \sep
Security \sep
Radiation Detection\sep
Pulse-Height Sharing
\end{keyword}

\end{frontmatter}






\section*{Acknowledgements}
This is NRCan/ESS Contribution 20140132.

\section{Introduction}

A portable Compton gamma-imaging device is a helpful tool in the prevention or clean-up of a radiological or nuclear incident. First responders require portable radiation detectors operating in the 0 - 3 MeV range that can be deployed in the field. To meet their needs, we have developed a Compton imager that is low-cost, ruggedize-able, and adaptable to suit a number of different surveying platforms (truck, helicopter, etc.). The detector is highly sensitive, (capable of localizing a 10~mCi~\Cs~source at 40~m distance with degree-level precision in 60~s), making it suitable for real-life situations involving radiation sources that can be weak, distant or shielded.

Some highly-compact Compton-imaging designs based on scintillator~\cite{Kataoka2013403} and semiconductor technologies (CZT~\cite{h3dpatent,CZTRedlen})) have been developed for commercial use. In the future, development in crystal growth technology may make these viable options for affordable larger-scale detectors. However, at present, these small ultra-portable detectors have limited sensitivity. In parallel, extremely large highly-sensitive imagers have been developed ~\cite{mitchell2009mobile,sords,sordspatent,ziock2008fieldable}. However, these do not meet the requirements of agencies with limited funds, nor are they suitable for mounting onto multiple platforms. Our detector aims to operate in the middle range, by being both sensitive and portable (can be loaded into and out of a vehicle), while remaining affordable. A more expensive design based on scintillator with SiPMs has been developed~\cite{Saullsipm2012}. The design presented in this paper employs long bars of scintillator (NaI(Tl)) with photomultiplier tube (PMT) readout, a combination well-proven for application in robust low-cost detectors operating in the field. 

\begin{figure}[!h]
\centering
\includegraphics[width=0.35\textwidth]{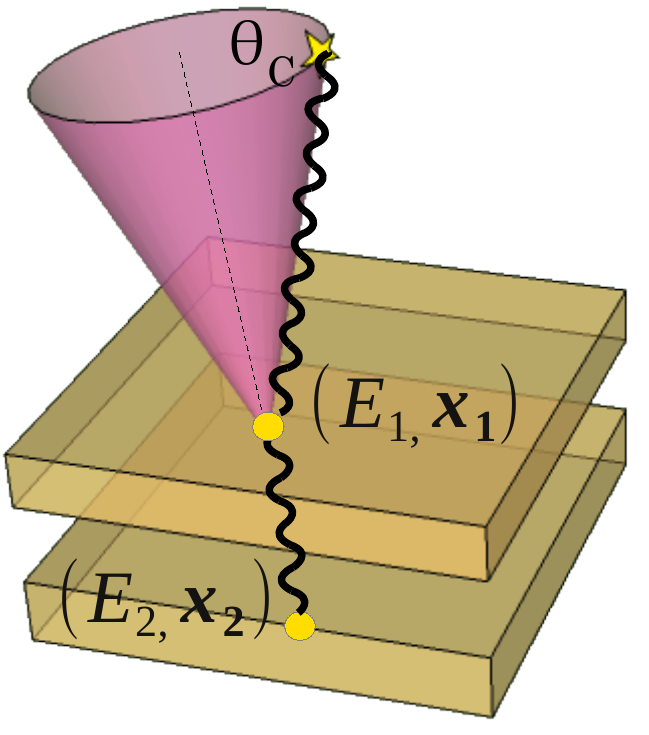} 
\caption{Compton imager schematic depicting the energy and position of the scattered electron, $E_1$, {\bf{\emph{x}}$\bf{_1}$}, recorded in the scatter plane, and the energy and position of the scattered gamma ray, $E_2$ and {\bf{\emph{x}}$\bf{_2}$}, recorded in the absorber plane. The Compton cone, of opening angle $\theta_C$, is calculated using the energy deposits and localizes the possible source locations.}
\label{fig:comptoncones}
\end{figure}

\section{Compton Imaging}

In a Compton gamma imager, an incoming gamma ray of energy $E_\gamma$ Compton interacts in a scatter detector, depositing energy $E_1$. The outgoing scattered gamma ray deposits its energy $E_2$ in an absorber detector. Neglecting Doppler broadening, the scattering angle between the initial and final-state gamma rays, $\theta_C$, can be determined from the two energy deposits, via
 \begin{equation}
 \cos\theta_C{} = 1 + m_0c^2(\frac{1}{E_\gamma} - \frac{1}{E_2}),
 \end{equation}
 where $E_\gamma = E_1 + E_2$ and $m_0c^2$ is the electron rest energy.
 As illustrated in Figure~\ref{fig:comptoncones}, the position of the
 source lies somewhere on a cone of opening
 angle $\theta_C$, where the cone axis is on the line between the two energy deposits. The location of the source can be reconstructed from the intersection of several cones.

\section{Bar Design}

We have designed, constructed and tested a Compton imager made of long bars of NaI(Tl) read out by PMTs fixed to the bar ends. Pulse-height sharing between the PMT signals is used to determine the position of the interaction. The use of long bars of NaI(Tl) offers an inexpensive design solution, as NaI(Tl) is relatively low cost, and the bar format requires few readout channels.

Long scintillation bars have previously been employed in Compton imagers where the energy deposits are large~\cite{sords}, however to employ this technique at lower energies, an optimization of the attenuation length of the bars is needed. Studies have been done to look at the effect of tuning the attenuation length of scintillator in order to optimize its position reconstruction~\cite{shao,huber, carter, labanti}, which holds well at high incident gamma-ray energies where one is not photo-statistics limited. However, operating at low energies (less than 150~keV in the scatter layer for incoming 662~keV gamma rays), requires tuning the bar attenuation length to optimize the overall performance of the imager. To this end, a highly detailed Monte-Carlo simulation using optical transport was developed to determine the optimal attenuation length of the scintillator bars. A prescription was then developed to manufacture these bars for the detector.

\FloatBarrier
\section{The Detector}
\label{sec:detector}

\begin{figure}[!h]
\centering
\includegraphics[width=0.45\textwidth]{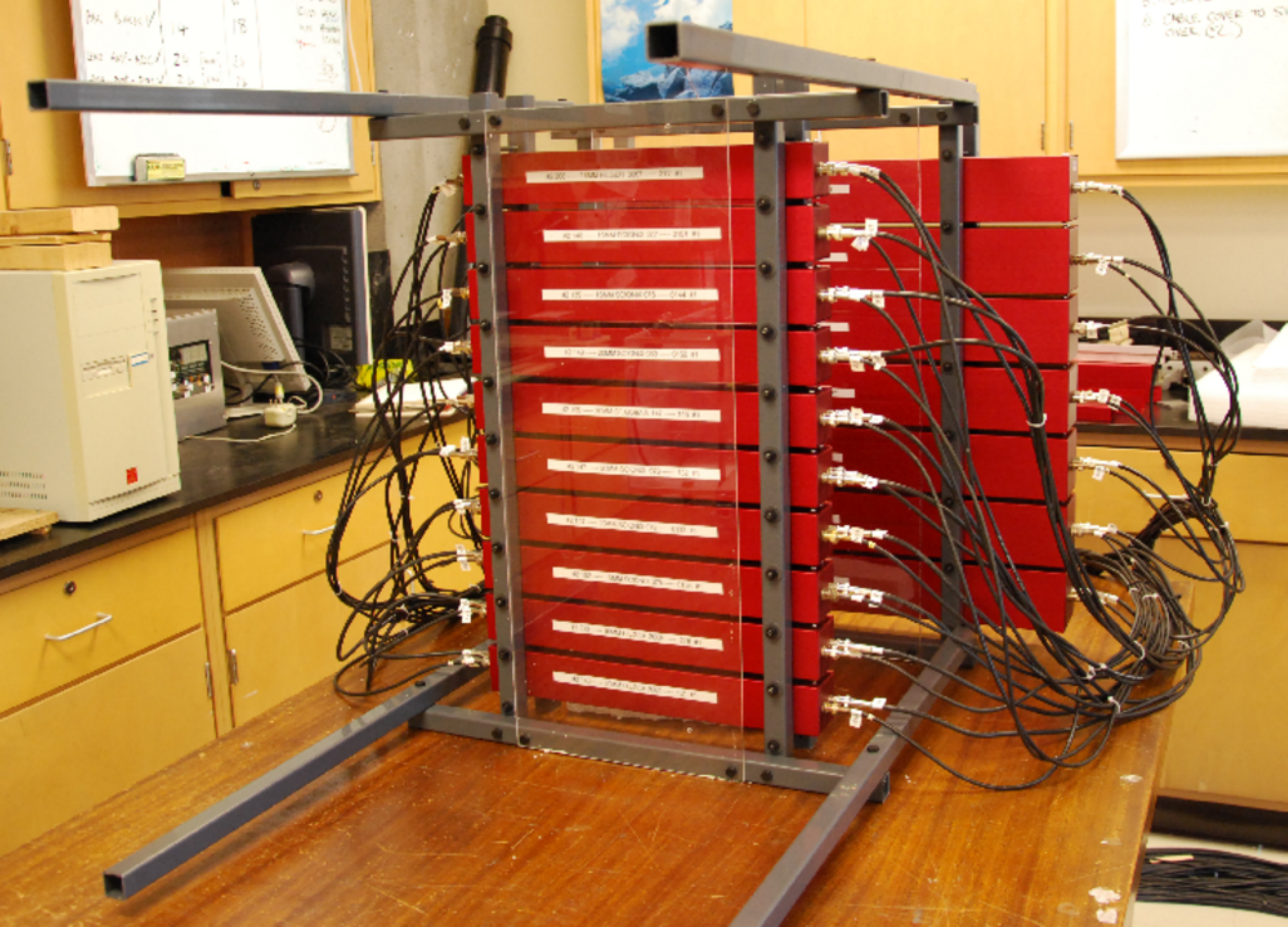}
\caption{Prototype consisting of ten scatter modules (front) and seven absorber modules (back) with a scatter-absorber spacing of 425~mm.}
\label{fig:prototype}
\end{figure}

\begin{figure}[!h]
\centering
\includegraphics[width=0.9\textwidth]{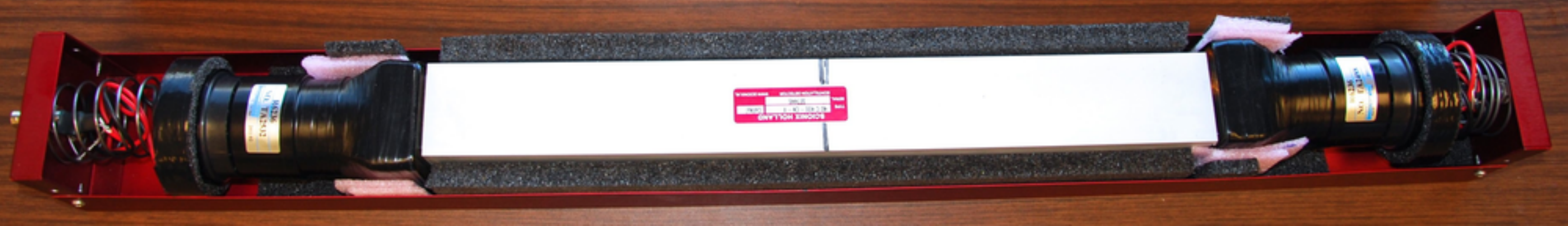}
\caption{A single absorber detector module. The scintillation light from a bar of encapsulated NaI(Tl) is read out by PMTs placed at the ends of the bar. The bar and PMTs are enclosed in a thin aluminum shell and packed with foam for protection.}
\label{fig:singleBar}
\end{figure}

\begin{table}
\centering
\renewcommand{\arraystretch}{1.3}
\setlength{\tabcolsep}{4pt}
\begin{tabular}{|c|c|c|c|c|c|}
\hline
\textbf{Plane} & \multicolumn{2}{|c|}{ \textbf{Scintillator}}  & \multicolumn{3}{|c|}{\textbf{PMT}} \\ 
 \cline{2-6}
 & number & size (mm$^3$)  &  model &  size (mm)  & QE\\  \hline
& 5 & 20~x~20~x~200  & R8900U-100 & 23.5  & SBA (35\%) \\ \cline{2-6}
\textbf{Scatter}  & 4 & 16~x~16~x~200  & R8900U-100 & 23.5  & SBA (35\%) \\ \cline{2-6}
  & 1 & 16~x~16~x~200  & R7600U-200 & 18.0 & UBA (43\%) \\ \hline
\textbf{Absorber} & 7 & 40~x~40~x~400  & R6236 & 54.0  & 23\% \\ 
\hline
\end{tabular}
\caption{Summary of the detector components making up the scatter and absorber plane. Only four of the seven absorber bars were included in the detector for assessing the performance for reasons discussed in Section~\ref{sec:barfail}.}
\label{table:readout}
\end{table}

 Previous work has shown that for the scatter plane, the optimal thickness for one Compton scatter of 662~keV gamma rays in NaI(Tl) is $\sim$~20~mm~\cite{sorma}. With this thickness, approximately 20\% of 662~keV gamma rays Compton scatter once and exit. For the absorber plane, a thickness of 40~mm is chosen, a compromise among increasing the absorption fraction, maintaining comparable lateral and longitudinal position resolution, and matching the bar cross section with commercially available PMTs.

 A photograph of the full-scale prototype is shown in Figure~\ref{fig:prototype}. The modular scatter and absorber layers are 425~mm  apart (center to center) and mounted on a frame constructed of 1/2'' PVC tubing. The scatter plane consists of ten \bara~and \barb~bars of NaI(Tl) and the absorber plane consists of seven \barc~bars of NaI(Tl).

An overview of the scintillator dimensions and the PMT sizes, QEs and model numbers of all the components are provided in Table~\ref{table:readout}. In the scatterer, we used both Super bialkali (SBA) PMTs, with QE~$\sim$~35\%, and Ultra bialkali (UBA) PMTs, with QE~$\sim$~43\%. In the absorber, we used standard PMTs. The scintillator was purchased from Scionix, Saint Gobain and Hilger Crystals\footnote[2]{Scionix: www.scionix.nl; Saint Gobain: www.saint-gobain.com; Hilger Crystals: www.hilger-crystals.co.uk.}, while the PMTs were from Hamamatsu\footnote[3]{www.hamamatsu.com}. 

 Each NaI(Tl) bar is encapsulated in aluminum with a quartz window located at each end. The bars are read out by PMTs glued to the windows. The entire setup is positioned inside a 1/16'' thickness aluminum container (see Figure~\ref{fig:singleBar}) and packed with foam to protect the assemblies from mechanical shock.  The PMT high voltage (HV) is supplied by a CAEN\footnote[1]{www.caen.it} SY 2527 Universal Multichannel Power Supply System.

\label{sec:trigger}
The data acquisition uses a combination of VME, CAMAC and NIM standards. The PMT pulses are amplified by a factor of ten (Phillips Scientific\footnote[4]{http://www.phillipsscientific.com} 776) and split into two sets. The first set goes into a discriminator (Phillips Scientific 705) and participates in a hardware coincidence effectively requiring one or more end-to-end coincidences in the scatter layer together with the same pattern in the absorber layer (Phillips Scientific 754 Logic Units). The second set of amplified PMT signals is fed directly into a 48-channel 10-bit VME VF48 digitizer~\cite{vf48} and integrated on-board when a coincidence is present. The resulting charge values are then transferred to a computer via a USB-VME bridge (CAEN V1718). Immediately, a software cut is made to include only events with exactly one scatter bar hit and one absorber hit, where a hit denotes that both the left and right PMTs of a given bar are above threshold. The energy deposit is reconstructed in a given bar by summing the signals of the left PMT~(S1) and right PMT~(S2). The position is reconstructed using the ratio of these two signals, S1/S2.

\FloatBarrier
\section{Development and Performance of Scintillator Bars}

\subsection{Motivation}

The photon propagation in a scintillator is determined by its physical and optical properties as well as its surface finish, coating and wrapping. The key to our design is tuning the attenuation length of the scintillator bars in order to improve the position resolution while maintaining as good an energy resolution as possible. The attenuation can be tuned by modifying the surface finish, e.g. polishing or degrading the smoothness~\cite{huber, shao,labanti}. This section provides a detailed explanation of how the energy and position reconstruction depend on the attenuation. The optical absorption coefficient, $\alpha$, multiplied by the length of the bar, $L$, is used here as a measure of the attenuation. 

\begin{figure}[!h]              
\centering
\includegraphics[width=0.45\textwidth]{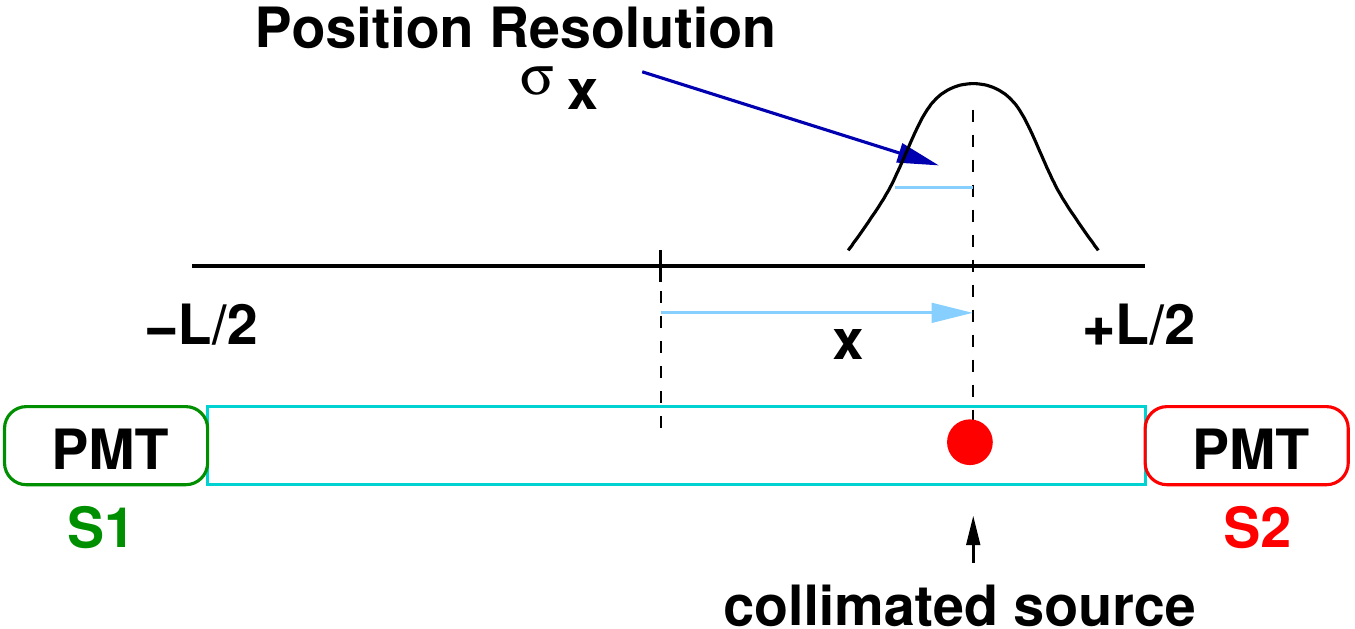}
\caption{Schematic of the position reconstruction using a bar of length L indicating $S_1$ and $S_2$, the signals from the left and right PMTs, and $x$, the displacement of the interaction from the bar center.}                                                           
\label{fig:positionreconstruction}

\end{figure}    

\noindent

\subsection{Position Resolution vs. Attenuation Length}

For an interaction at the longitudinal position $x$ in a bar of length $L$, the signal $S_{1}$ from the left PMT can be written as

\begin{equation}
S_{1} = \frac{N_{0}}{2}Pe^{-\alpha(\frac{L}{2} + x)} ,
\label{eq:s1}
\end{equation}

\noindent
and the signal from the right PMT as

\begin{equation}
S_{2} = \frac{N_{0}}{2}Pe^{-\alpha(\frac{L}{2} - x)} ,
\label{eq:s2}
\end{equation}

\noindent where $N_{0}$ is the number of photons generated and $P$ is the probability of detecting each photon. The probability, $P$, incorporates all sources of photon loss that are independent of the position of interaction, such as the probability of generating a photon, transmission at the scintillator ends, and PMT QE. $L$ and $x$ are the scintillator length and the displacement of the interaction from the bar center, respectively (see Figure~\ref{fig:positionreconstruction}).

Propagating the photostatistical uncertainties of $S_1$ and $S_2$, yields an expression for the position resolution~\cite{carter}:

\begin{equation}
\sigma_{X \mbox{\tiny{Stat}}} = \frac{1}{\alpha\sqrt{N_{0}P}} e^{\frac{\alpha L}{4}}\sqrt{\cosh(\alpha x)}.
\label{eq:deltax2}
\end{equation}

\noindent
Equation~\ref{eq:deltax2} highlights what is known to be true empirically. The position resolution depends on the initial number of photons produced, $N_{0}$, the probability of detecting each photon, $P$, and the absorption coefficient, $\alpha$. The minimum of the position resolution along a bar of length $L$ occurs when $\alpha L$ = 2.9~\cite{carter}. $\alpha L$ for a polished bar of NaI is much less ($\alpha L$~$\sim$~0.3 for a \barb~bar). Thus, increasing $\alpha$ should improve the position resolution. However, increasing $\alpha$ also has the undesirable effect of increasing the position dependence of the position resolution.

\subsection{Energy Resolution vs. Attenuation Length}

The expression $\sqrt{S_1 \times S_2}$ is proportional to $N_0$ and independent of $x$ and can thus be used to reconstruct the energy without knowing the position where deposition occurred. A derivation similar to the position resolution calculation can be used to obtain an expression for the relative uncertainty of the energy due to statistical fluctuations by propagating the statistical uncertainties of $S_1$ and $S_2$:

\begin{equation}
\frac{\sigma_{E \mbox{\tiny{Stat}}}}{E}=\frac{1}{\sqrt{N_{0}P}} e^{\frac{\alpha L}{4}}\sqrt{\cosh(\alpha x)}.
\label{eq:s1s2red3}
\end{equation}

Equation~\ref{eq:s1s2red3} shows that increasing the number of detected photons, either by increasing $N_0$ or $P$, improves the energy resolution whereas increasing $\alpha$ worsens the energy resolution. Increasing $\alpha$ also has the undesirable effect of increasing the position dependence of the energy resolution. The situation is therefore complicated. Decreasing the attenuation length (increasing $\alpha L$) both improves the position resolution and worsens the energy resolution. Thus, to determine the ideal attenuation length, an optimization is undertaken with simulations. 

\subsubsection{Relating the Energy Resolution to the Position Resolution}
\label{prescalc}

The total energy resolution, $\sigma_{E}$, is equal to a combination of $\sigma_{E  \mbox{\tiny{Stat}}}$, the energy resolution due to statistical fluctuations, and $\sigma_{E \mbox{\tiny{Int}}}$, the intrinsic energy resolution,

\begin{equation}
\sigma_{E}^2 = {\sigma_{E \mbox{\tiny{Stat}}}}^2 + {\sigma_{E \mbox{\tiny{Int}}}}^2.
\label{eq:esum}
\end{equation}

By combining Equations~\ref{eq:deltax2} and~\ref{eq:s1s2red3}, one obtains an expression relating the statistical component of the position resolution to the statistical component of the energy resolution,

\begin{equation}
\sigma_{X \mbox{\tiny{Stat}}} = \frac{1}{\alpha} \frac{\sigma_{E \mbox{\tiny{Stat}}}}{E}.
\label{eq:XvsE}
\end{equation}

 Equation~\ref{eq:XvsE} is then combined with Equation~\ref{eq:esum} to obtain an expression for the position resolution, $\sigma_{X}$, as a function of the measured total energy resolution, ($\frac{\sigma_{E}}{E}$),

\begin{equation}
\sigma_{X} \approx \sigma_{X \mbox{\tiny{Stat}}} = \frac{1}{\alpha} \sqrt{ \left( {\frac{\sigma_{E}}{E}} \right)^2 -  \left( {\frac{\sigma_{E  \mbox{\tiny{Int}}}}{E}}  \right)^2} .
\label{eq:preseres}
\end{equation}

By assuming a fixed value of 6\% for the intrinsic energy uncertainty, we use Equation~\ref{eq:preseres} to obtain an estimate of the position resolution indirectly from the energy resolution measurement at energies for which the position resolution is not directly measured. Note that the intrinsic energy resolution depends on the quality of the crystal and is not measured for our scintillator. Thus, the assumption of a fixed 6\% intrinsic energy uncertainty contributes to a moderate error in the estimation of the position uncertainty at higher energies ($\sim$ 5\% error at 662~keV), with the error decreasing at low energies where the uncertainties from photostatistics dominate.

\subsection{Optimal $\alpha$L from Simulations}
\label{sec:alphaL}
GEANT4~\cite{geant4toolkit} simulations were used to determine the optimal attenuation length of the bars. 

First, a detailed optical-photon transport was carried out to determine the energy and position resolution for NaI(Tl) bars as a function of attenuation length. Different attenuation lengths were modeled by making part of the bar surface absorbent. The surface treatment is referred to as {\it polished front-painted} in GEANT4 and is specularly reflective with 96\% reflectivity. The bars simulated were of sizes \barb~and \barc~and had 0\%, 25\%, 50\%, 75\% and 100\% of one side surface-treated, exploring $\alpha L$ values in the range of $\sim$~0.3 to 2.5. A monoenergetic 662~keV gamma-ray line source was directed towards the bar at multiple positions along its length, and the reconstructed energy and position resolution values were used to extract resolution functions as input for further modeling.

Simulations were then carried out for a complete detector made up of ten 20~mm-width scatter bars and seven 40~mm-width absorber bars, with a vertical spacing between the bars of 65~mm for the scatterer and 90~mm for the absorber. The separation between the scatter and absorber plane was fixed at 425~mm. The performance of 25 detectors corresponding to the different combinations (5 $\times$ 5) of scatterer and absorber were compared. For each detector combination, a simulation equivalent to a three-hour exposure to a 10~mCi \Cs~source located on the detector symmetry axis 40~m away was carried out. The energy and position of the interactions recorded in GEANT4 were smeared using the energy and position resolution functions derived from the detailed optical simulation. Only events with full gamma-ray energy deposit shared between exactly one scatter bar and exactly one absorber bar were selected. Furthermore, events exhibiting the energy signature of a back-scatter event (incoming gamma scattering $\sim$ 180$^{\circ}$ backwards from absorber to scatterer) were rejected.

\begin{figure}[!h]
\begin{tabular}{cc}
\centering
\includegraphics[width=0.45\textwidth]{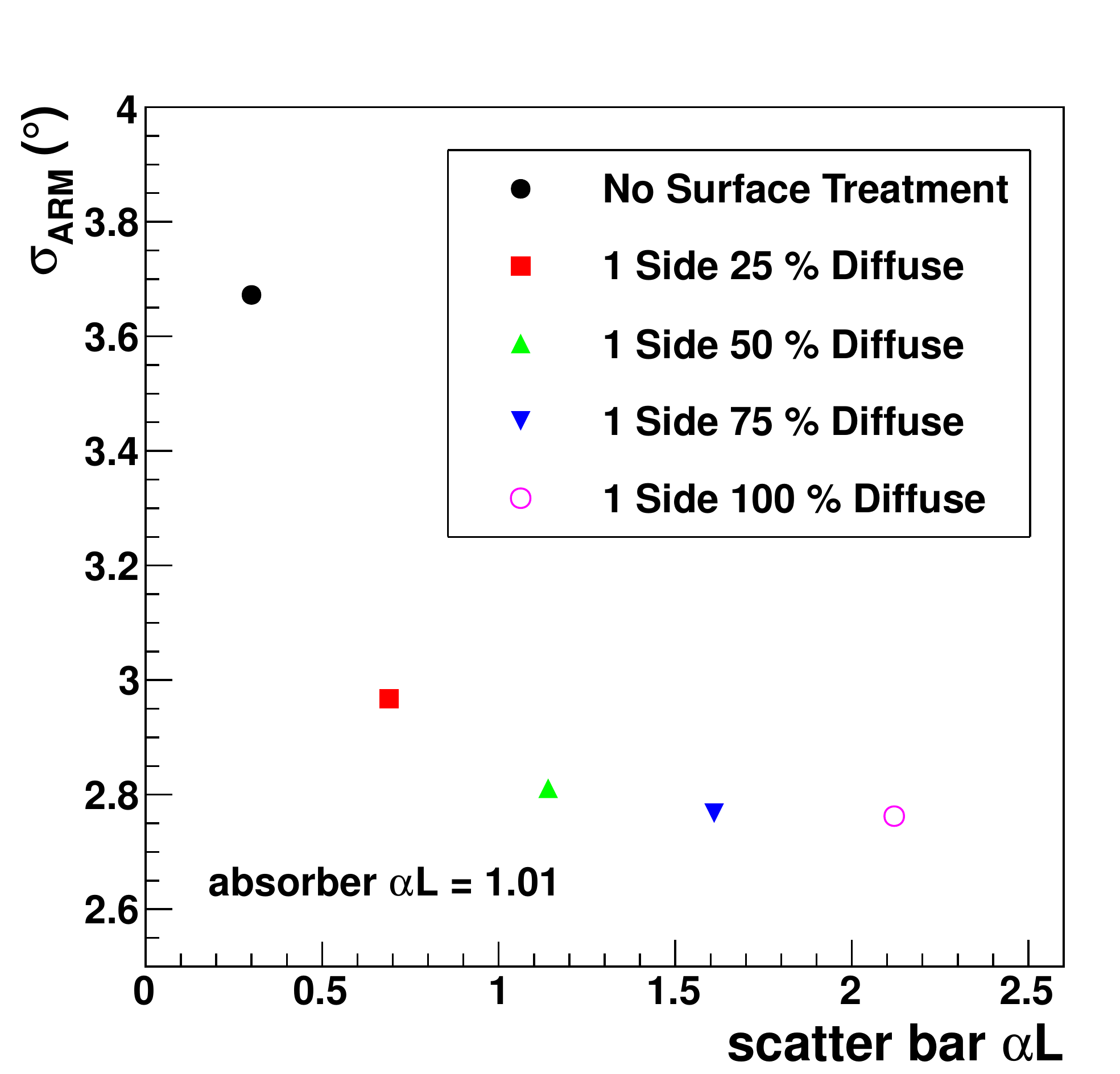} & 
\includegraphics[width=0.45\textwidth]{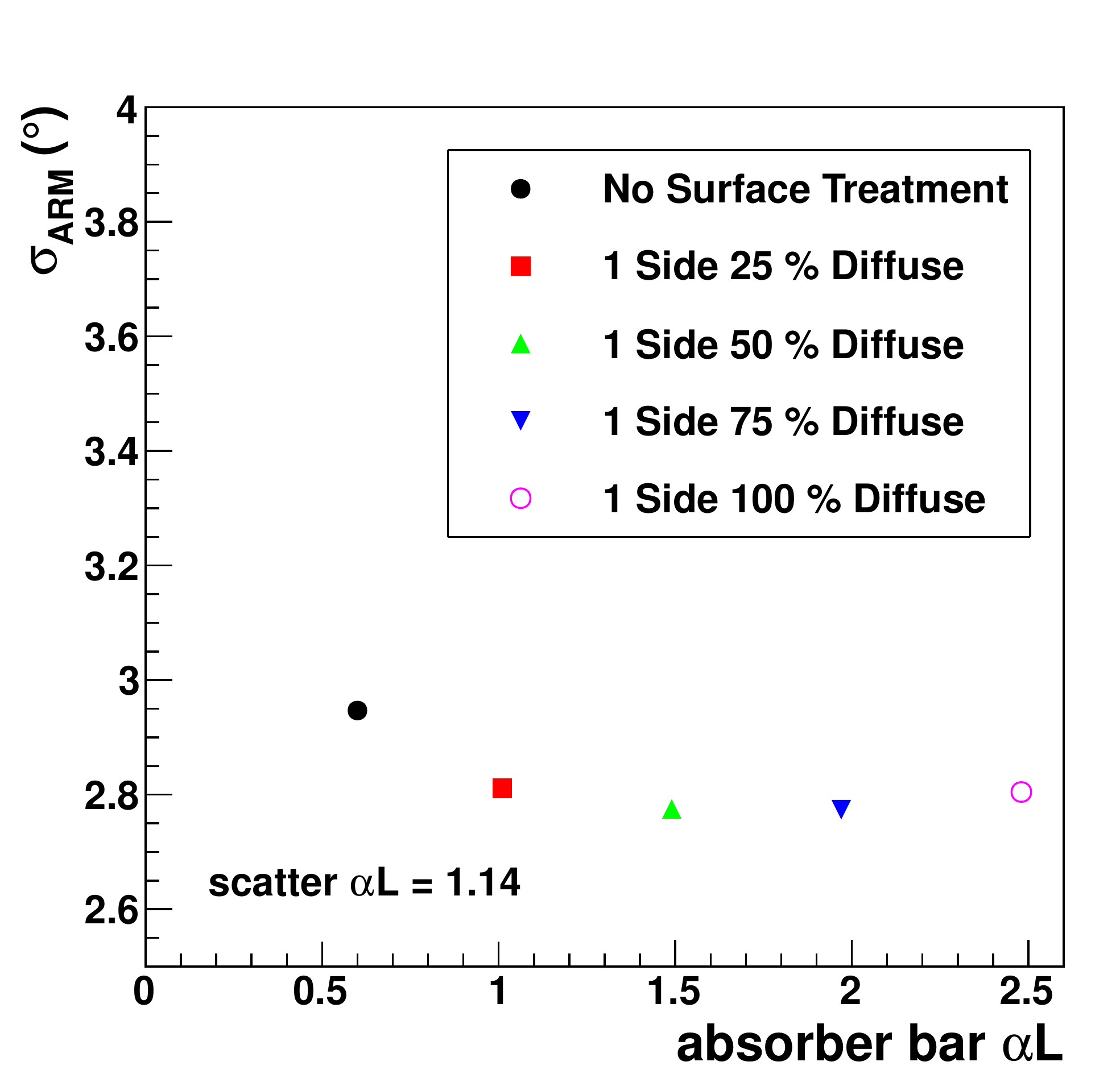}  
\end{tabular}
\caption{Simulation. Left: \sigARM~as a function of scatter bar $\alpha L$, for absorber $\alpha L$ fixed at 1.01 (25\% of one side surface-treated). Right: \sigARM~as a function of absorber bar $\alpha L$, for scatter $\alpha L$ fixed at 1.14 (50\% of one side surface-treated). Note that the y axis is zero-suppressed and the error bars are smaller than the symbols in both the left and right figures.}
\label{fig:ARMFSvsAlpha}

\end{figure}

The width of the angular resolution measure (ARM) for the selected events was used as a figure of merit for detector performance. ARM$~=~\theta_C - \theta_{\mbox{\scriptsize geom}}$, where $\theta_C$ is the Compton angle calculated from the energy deposits and $\theta_{\mbox{\scriptsize geom}}$ is the observed scattering angle based on the known source position and the reconstructed hit positions. Figure~\ref{fig:ARMFSvsAlpha} (left) shows the standard deviation, \sigARM, as a function of scatter bar $\alpha L$, with the absorber $\alpha L$ fixed at 1.01 (25\% of one side surface-treated). A significant improvement ($\sim$~26\%) in \sigARM~is observed, decreasing from 3.7$^{\circ}$ to 2.8$^{\circ}$ with increasing surface treatment. Figure~\ref{fig:ARMFSvsAlpha} (right) shows \sigARM~as a function of absorber bar $\alpha L$, with the scatter $\alpha L$ fixed at 1.14 (50\% of one side surface-treated). Once again, \sigARM~decreases with increasing $\alpha L$, leveling off at $\alpha L$~$\sim$~1. The ARM for the remaining possible combinations of scatter and absorber $\alpha L$ demonstrated similar behaviour. Since decreasing the attenuation length (or increasing $\alpha$) also increases the non-linearity of the energy and position resolution along the bar, an $\alpha L$ of 1, where the improvement in \sigARM~levels off, was deemed optimal.

\FloatBarrier

\subsection{Determination of the Surface Treatment}

\begin{figure}[!h]
\centering
\resizebox{0.45\textwidth}{!}{\includegraphics{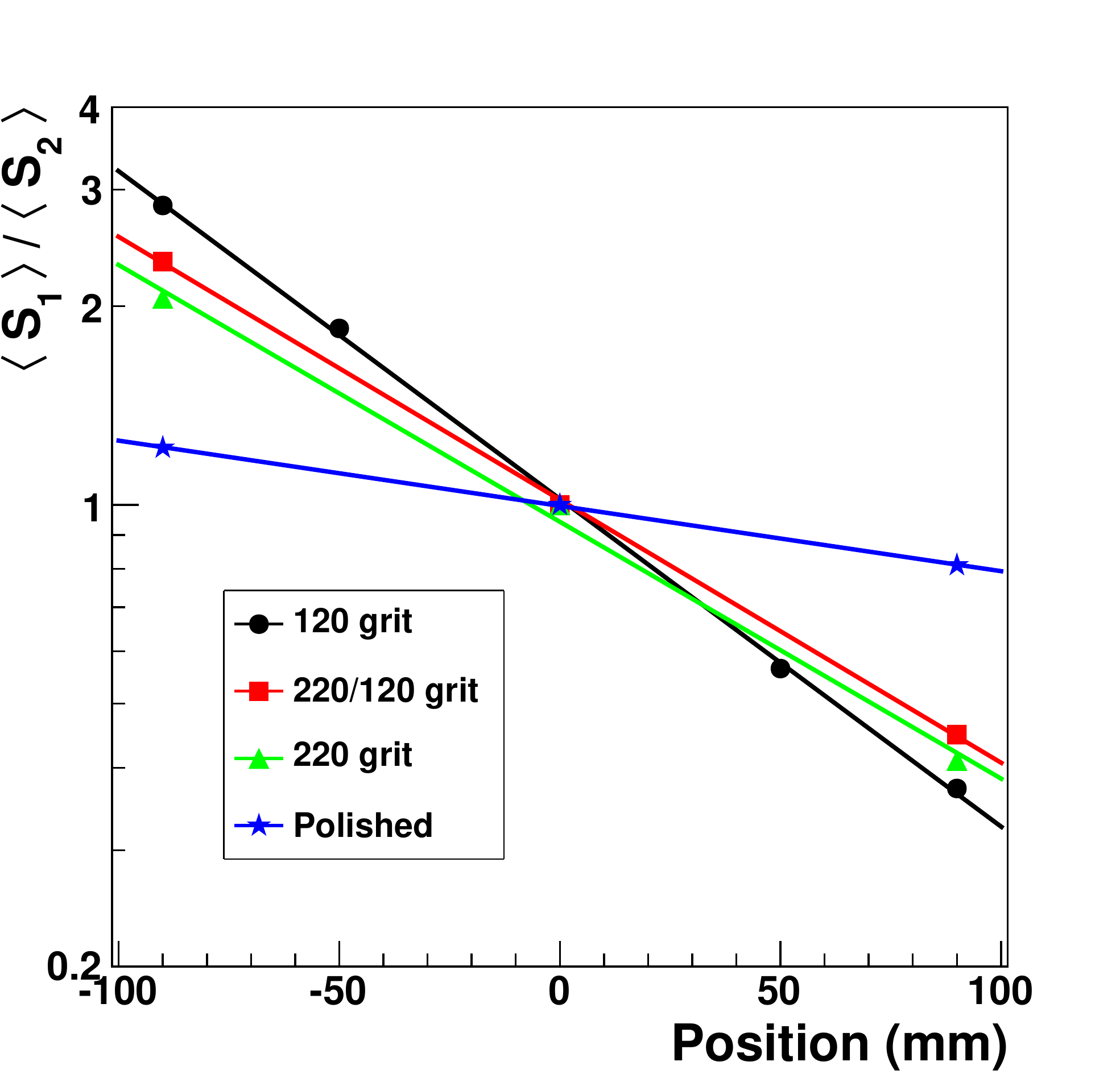}}
\caption{Plot of the ratio, \sratio~for a \barb~bar at 662~keV, as a function of position for a polished bar (blue stars); four sides degraded with 220 grit sandpaper (green triangles); two sides degraded with 220 grit and the remaining two sides with 120 grit (red squares); and four sides degraded with 120 grit (black circles). The data sets can be described by exponential functions. Note that the data have been scaled such that the middle point is located at (0,1).}
\label{fig:ratioScionix}
\end{figure}

\begin{table}[!h]
\centering
\renewcommand{\arraystretch}{1.6}                                                                                            
\setlength{\tabcolsep}{7pt}                                                                                                 
\begin{tabular}{|c|c|c|c|}
\hline
\centering
{ Surface Treatment } & { $\alpha$ L}  & { $\frac{\Delta E}{E}$}  & { $\sigma_x$} \\
{ }  &  & { (\% FWHM)} & { (mm) } \\ \hline
\hline
Polished & 0.23 & 7.3 & 18 \\
220 & 0.90 & 8.4 & 7 \\
220/120 & 0.92 & 9.8 & 8\\
120 & 1.14 & 9.5 & 6\\
\hline
\end{tabular}
\caption{Scionix data for $\alpha L$ measured for a~20~mm bar with successive levels of sanding and the energy resolution measured at the bar center at 662~keV. The position resolution values presented here are estimated from the energy resolution and $\alpha$ using the technique described in Section~\ref{prescalc}.}
\label{table:alpha LScionix20}
\end{table}

 A technique to modify the bar surface was developed with the manufacturer Scionix to achieve scintillator bars with attenuation lengths of $\alpha L$ $\sim$ 1. Due to the hygroscopic nature of NaI(Tl), any alteration of the scintillator surface must take place in a low-humidity environment. Thus, the procedure to modify the bar surface was undertaken in a dry room at the manufacturer's site and carried out by the manufacturer.

To tune the attenuation length, the bar surface was degraded by sanding with various grit sizes in the longitudinal direction. Four different surfaces were compared: polished; four sides degraded with 220 grit sandpaper; two sides degraded with 220 grit and the remaining two sides with 120 grit; and four sides degraded with 120 grit.
Figure~\ref{fig:ratioScionix} shows the ratio of the mean values of the signals at the two ends, \sratio, versus the position of a collimated 662~keV energy source along the bar, for these four surface treatments.
As the sandpaper grit number decreases (becomes more abrasive), the attenuation of the bar increases.
The attenuation length was obtained by applying an exponential fit to the data. As shown in Table~\ref{table:alpha LScionix20}, $\alpha L$ increases from 0.23 for an untreated bar to 1.14 for 120 grit applied to all four sides.

Energy resolutions measured with the 662~keV source at the center of the treated bars are also presented in Table~\ref{table:alpha LScionix20}, along with position resolutions calculated using the technique described in Section~\ref{prescalc}. (Note that the ratio values used to determine $\alpha L$ and the energy resolutions were provided by Scionix without uncertainties.)

We find, as expected, a dramatic improvement in position resolution as $\alpha L$ increases to close to 1 with only minor improvements with further degradation of the bar surface. The accompanying worsening of energy resolution is comparatively minor.  The second most attenuating surface treatment (220 grit to two sides and 120 grit to the two other sides) was chosen for the 20~mm bars, yielding $\alpha L$ = 0.92, which is close to the targeted value of 1.

The same procedure was then repeated for the 16~mm and 40~mm bars to obtain $\alpha L$ values $\sim$ 1.  For the 16 mm bar, the same combination of 220~grit and 120~grit was chosen, while for the 40~mm bar, a surface treatment of 120 grit was chosen. To test the reproducibility of the technique, bars were then ordered from two other manufacturers, Hilger and Saint-Gobain, who employed the same prescription for the surface treatment.

\FloatBarrier

\subsection{Performance of the Bars}
\label{sec: barperformance}

In total, 18 bars were tested from three manufacturers, Scionix, Hilger, and Saint-Gobain. The bars came in three sizes: 16~mm width (five bars), 20~mm width (six bars), and 40~mm width (seven bars). Each set had the same surface treatment applied to it. For in-house testing, the bars were paired with PMTs having areas that matched their cross section. The 16~mm and 20~mm bars were paired with R8900U-100 PMTs (QE~=~35\%) while the 40~mm bars were paired with R6236 PMTs (QE~=~23\%). The bars were tested in their aluminium containers described in Section~\ref{sec:detector}.

We made a detailed measurement of the performance of the bars at a single energy, 662~keV. We parametrized the position as a function of the left and right PMT signals to reconstruct the position of the interactions. A 10~$\mu$Ci~\Cs~source was placed behind a cylindrical collimator having a thickness of 70~mm and an aperture of 5.5~mm. The collimator was attached to a motorized platform (Arrick Robotics\footnote[5]{www.arrickrobotics.com} Model XY-30).  Four scans of each bar were made, two in each direction, with 5~mm steps. The left and right PMT outputs ($S_1$ and $S_2$) were obtained by digitizing the PMT signals with a 1~GS/s, 8-bit digitizer (Acqiris\footnote[6]{www.agilent.com} DC270) and computing the area of each pulse. The digitizer was triggered by coincidence of the left and right PMT signals. At each step, 10~000 coincidence events were recorded. 

\begin{figure}[!h]
\centering
\begin{tabular}{ccc}
\resizebox{0.33\textwidth}{!}{\includegraphics{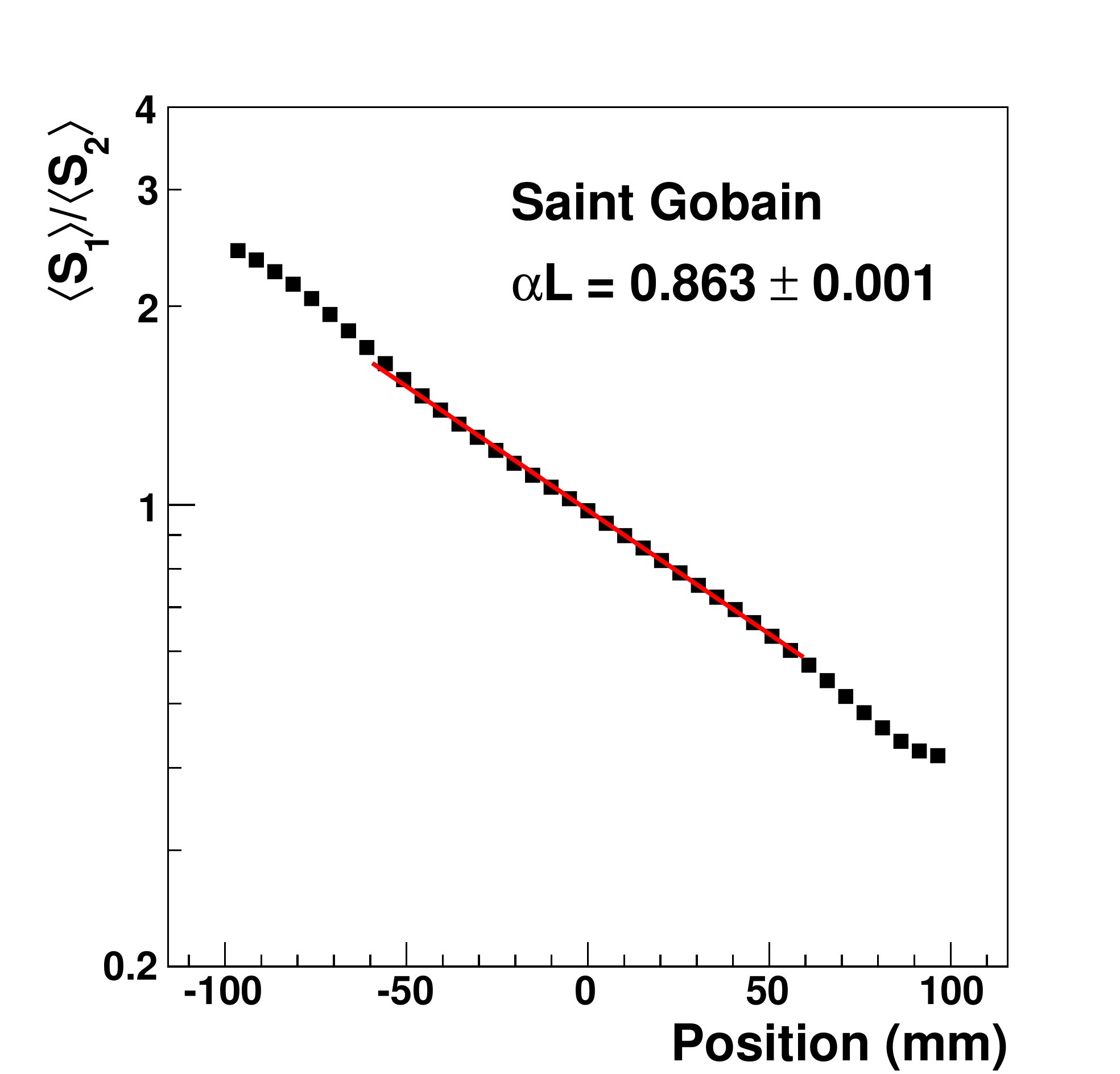}} &  
\resizebox{0.33\textwidth}{!}{\includegraphics{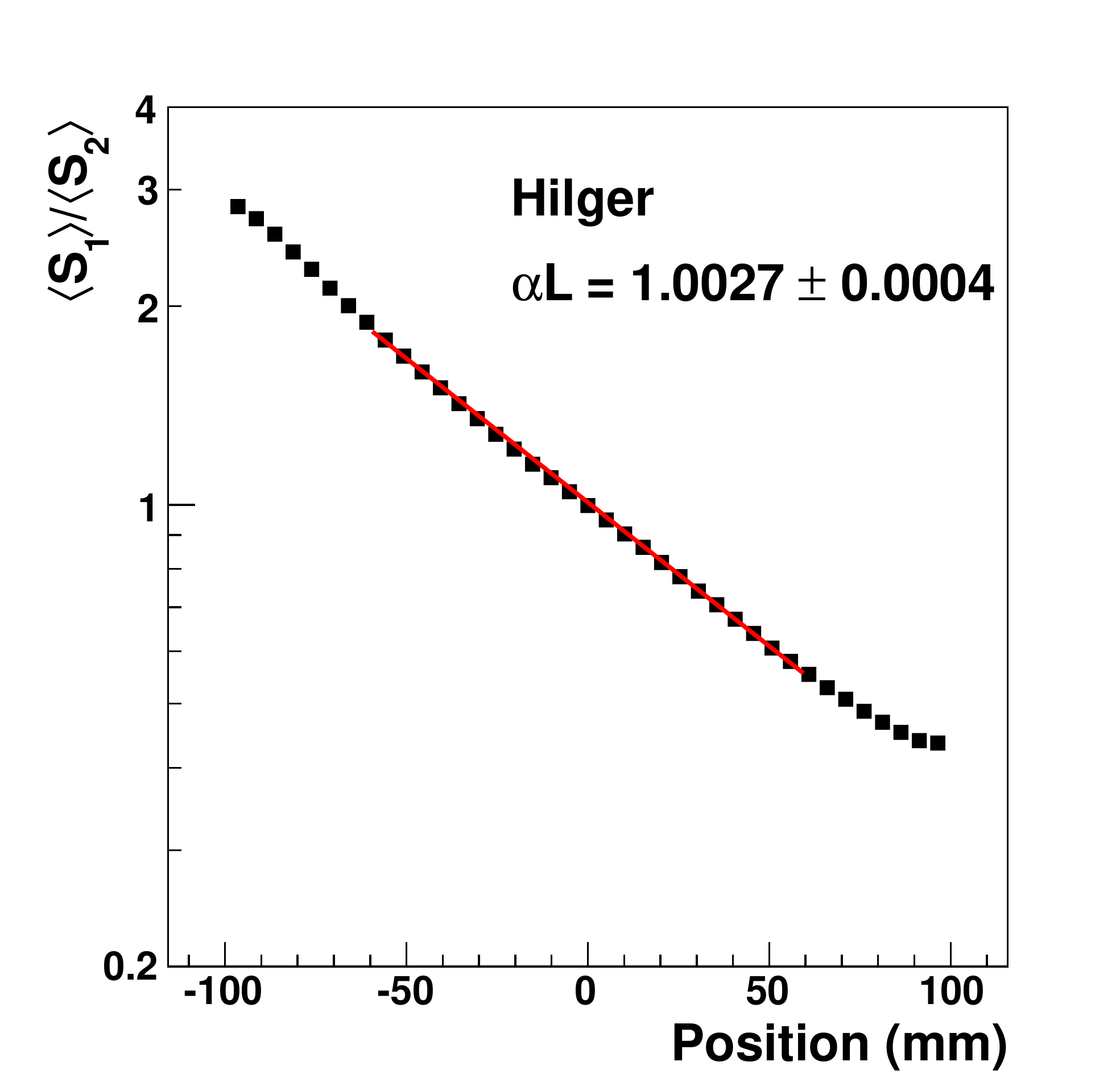}} & 
\resizebox{0.33\textwidth}{!}{\includegraphics{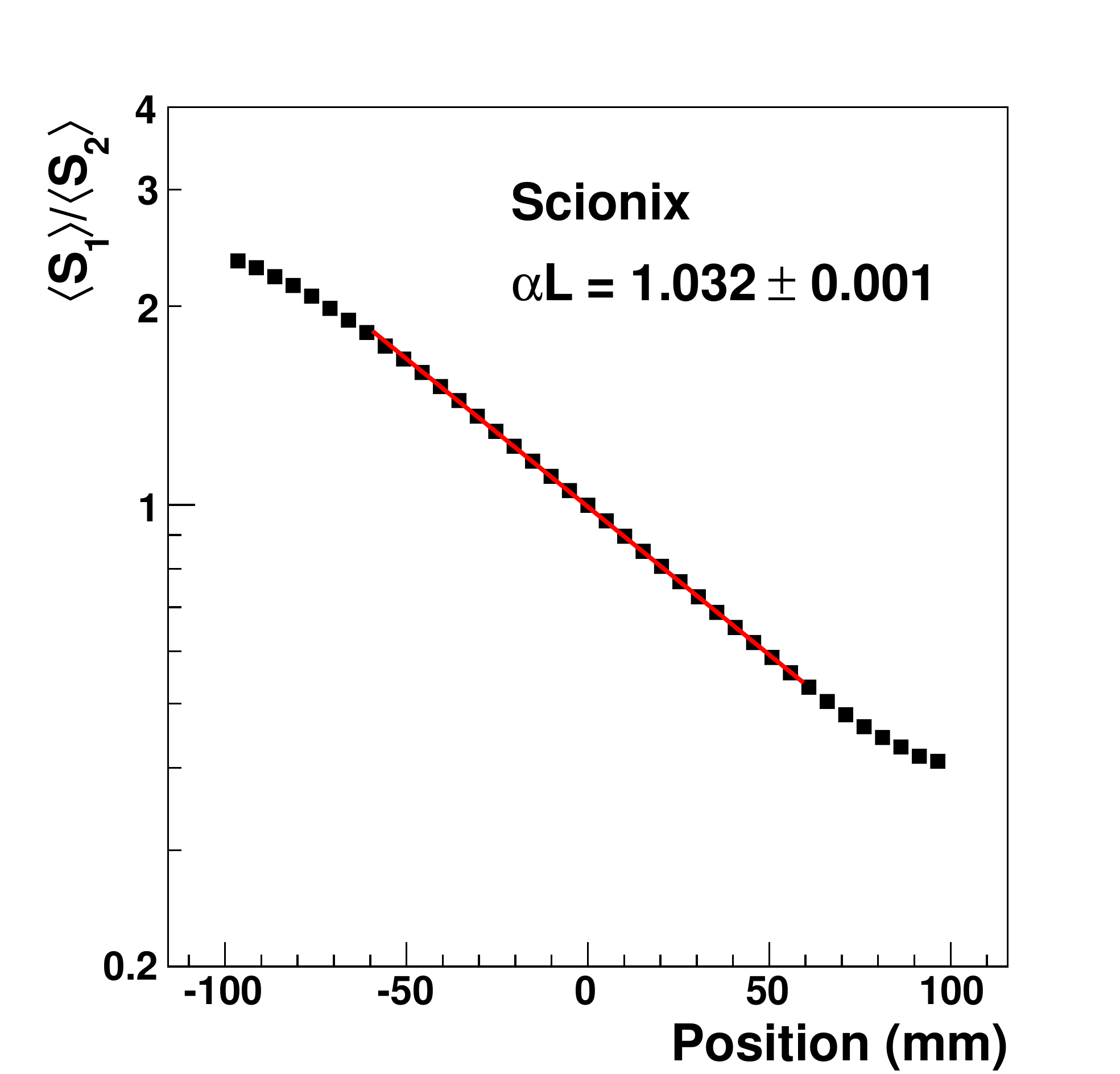}} \\ 
\end{tabular}
\caption{Plot of the ratio, $\sratioe$, for an individual 20~mm bar from three manufacturers (Saint-Gobain (left), and Hilger (middle) and Scionix (right)). To obtain $\alpha L$, the central region of each set of points is fit to an exponential function extending from - 60~mm to + 60~mm.}
\label{fig:ratioAll}
\end{figure}

To obtain the position response, Gaussians were fit to the photopeaks in $S_1$ and $S_2$ and the photopeak pulse-height means, $\avg{S_1}$ and $\avg{S_2}$, were measured. Figure~\ref{fig:ratioAll} shows the signal ratio,  $\sratioe$, versus distance along the bar, for an individual 20~mm bar from each of the three manufacturers.

To determine the position resolution, only events within 2.5 standard deviations of the mean of the photopeaks in $S_1$, $S_2$, and their sum, were considered. For these events, the ratio, \sratio, was calculated and converted to a position using the position versus signal characterization. The width of the resulting distribution of positions was adopted as the position resolution at 662~keV. 

Calibration of the bars at other energies was performed differently. Sources were placed only on the bar center. A Gaussian was fit to the summed signal to obtain a parametrization of the energy resolution and energy versus signal. To obtain a parametrization of the position resolutions for these energies, the approach described in Section~\ref{prescalc} was used.

The average $\alpha L$, and the energy and position resolutions obtained at 122~keV and 662~keV for the 16~mm and 20~mm scatter bars are shown in Tables~\ref{table:alphaL16ExpSim} and \ref{table:alphaL20ExpSim}, respectively. The uncertainties on the attenuation measurements (when $N$, the number of bars measured, was greater than one) were estimated from the spread between the different bars and their relatively small values ($\sim$~5\%) indicate that the application of the surface treatments is repeatable by a given manufacturer. The bars from the different manufacturers have similar attenuation. However, the measured energy and position resolutions differ significantly from one manufacturer to another, with the Hilger bars exhibiting the worst performance.

 The performance of the bars was also compared with the performance of a simulated bar of equivalent $\alpha L$. These results are shown in Table~\ref{table:alphaL16ExpSim} for the 16~mm bar, and similar results were observed for the other bar sizes. The measured energy and position resolution values at 662~keV were found to be in good agreement with the simulations. However,  those for 122~keV were found to be slightly worse than predicted (\lae 20\%), with the Hilger bars deviating the most from the simulations. The Hilger bars exhibited significantly worse energy and position resolution values. For these bars, we observed far fewer photoelectrons, which could help explain why both the energy and position resolution values are worse. This demonstrates the limitations in the simulations to model the optical photon transport of the bars of NaI(Tl) accurately, where the properties of the scintillator, scintillator surface, and PMT photocathode are difficult to model and also vary from one manufacturer to another.

The average $\alpha L$, and energy and position resolutions obtained at 662~keV for the 40~mm absorber bars are shown in Table~\ref{table:alphaL40ExpSim}. The Scionix bars are significantly more attenuating ($\alpha L$~=~1.10) in comparison to those from the other two manufacturers ($\alpha L$~$\sim$~0.8).

\begin{table}[!h]                                                                                                                
\centering
\renewcommand{\arraystretch}{1.6}                                                                                            
\setlength{\tabcolsep}{2pt}                                                                                                 
\begin{tabular}{|c|c|c|c|c|c|c|}
\hline
{Manufacturer} & {\Nbar} & {$\alpha L$} & {$\frac{\Delta E}{E}$} 122 keV & {$\frac{\Delta E}{E}$} 662 keV & {$\sigma _x$} 122 keV & {$\sigma _x$} 662 keV \\
 &   &   & {\% (FWHM)} & {\% (FWHM)}  & {(mm)} & {(mm)}\\ \hline
\hline
Hilger& 2  &1.02 $\pm$ 0.06 & 17.5~$\pm$~0.2 &8.1 $\pm$ 0.3 &  13.1 $\pm$ 0.9* & 5.6 $\pm$ 0.4 \\  
&  &  & (13.3 $\pm$ 0.3) & (8.4 $\pm$ 0.6) & (10.8 $\pm$ 0.1)  & (5.2 $\pm$ 0.3)\\ \hline
Scionix& 3 & 1.18 $\pm$ 0.01  & 15.4~$\pm$~0.4 & 8.2 $\pm$ 0.1 &  11.3 $\pm$ 0.2*  & 5.1 $\pm$ 0.1\\
 &  & & (13.7 $\pm$ 0.4) & (8.5 $\pm$ 0.7) & (9.6 $\pm$ 0.1)  & (4.9 $\pm$ 0.3) \\ \hline
\end{tabular}
\caption{$\alpha L$, and the energy resolution and position resolution measured at both 662~keV and 122~keV for the 16~mm bars provided by Hilger and Scionix. \Nbar represents the number of bars tested. The position resolution values followed by * are estimated from the energy resolution using the technique described in Section~\ref{prescalc}. The performance of a simulated bar of equivalent $\alpha L$ is presented in parentheses.} 
\label{table:alphaL16ExpSim}
\end{table}

\begin{table}[!h]                                                                                                                
\centering                                                                                                              
\renewcommand{\arraystretch}{1.6}                                                                                            
\setlength{\tabcolsep}{2pt}                                                                                                 
\begin{tabular}{|c|c|c|c|c|c|c|}
\hline
{Manufacturer}  & {\Nbar} & {$\alpha L$} & {$\frac{\Delta E}{E}$} 122 keV & {$\frac{\Delta E}{E}$} 662 keV & {$\sigma _x$} 122 keV & {$\sigma _x$} 662 keV \\
 & &   & {\% (FWHM)} & {\% (FWHM)}  & {(mm)} & {(mm)}\\ \hline
\hline
Saint-Gobain& 1 & 0.863 $\pm$ 0.001 & 13.8~$\pm$~0.1&8.0 $\pm$ 0.2& 12.1 $\pm$  0.1*  &5.7 $\pm$ 0.1 \\ 
Hilger& 2  & 0.999 $\pm$ 0.003 & 18.4~$\pm$~0.6 & 8.7 $\pm$ 0.2& 14.7 $\pm$ 0.5*  & 6.2 $\pm$ 0.3      \\  
Scionix& 3 & 1.03 $\pm$ 0.02 & 15.6~$\pm$~0.5 & 7.7 $\pm$ 0.3& 11.9 $\pm$ 0.6*  & 6.0 $\pm$ 0.1 \\
\hline
\end{tabular}
\caption{$\alpha L$, and the energy resolution and position resolution measured at 662~keV and 122~keV for the 20~mm bars. The position resolution values followed by * are estimated from the energy resolution using the technique described in Section~\ref{prescalc}.}
\label{table:alphaL20ExpSim}
\end{table} 

\begin{table}[!h]                                                                                                                
\centering                                                                                                              
\renewcommand{\arraystretch}{1.6}                                                                                            
\setlength{\tabcolsep}{4pt}                                                                                                 
\begin{tabular}{|c|c|c|c|c|}
\hline
{Manufacturer} & { \Nbar} & { $\alpha L$} & { $\frac{\Delta E}{E}$} 662 keV & { $\sigma _x$} 662 keV \\
 &  &   & { \% (FWHM)} & { (mm)}\\ \hline
\hline
Hilger    & 2 &0.76  $\pm$ 0.03  & 12.3 $\pm$ 0.9 &19.6 $\pm$ 2.7 \\
Saint-Gobain& 3 & 0.78 $\pm$ 0.02  & 8.2 $\pm$ 0.2  & 10.9 $\pm$ 0.9 \\ 
Scionix   & 2 &1.07  $\pm$ 0.08  & 7.9 $\pm$ 0.1  & 8.5 $\pm$ 0.1 \\
\hline
\end{tabular}
\caption{$\alpha L$, the energy resolution and position resolution measured at 662~keV for the 40~mm bars.}
\label{table:alphaL40ExpSim}
\end{table}

\FloatBarrier
\section{Bar Failures}
\label{sec:barfail}
Three of the seven absorber bars experienced a sudden change in performance. The attenuation length decreased, accompanied by an increase in the number of photoelectrons detected at one end of the bar and a decrease at the other end. Both the Scionix and Saint Gobain bars had this issue, but not the Hilger bars. This is currently under investigation and there are several possible hypotheses, such as contamination of the NaI crystal, problems with the glue on the glass-scintillator interface, failure of the plastic seal that surrounds the quartz windows at the bar ends, and cracking due to the thermal expansion of the crystal, glass and aluminum. Integrating the PMT into the encapsulation is a possible development that would improve ruggedness and may mitigate this problem.

\section{Detector Performance}

\begin{table}[!htp]
\centering
\renewcommand{\arraystretch}{1.3}
\setlength{\tabcolsep}{6pt}
\begin{tabular}{|c|c|c|}
\hline
 { Energy} & { Source} & {Source Strength}\\
  { (keV)} &  & { s$^{-1}$ ($\times 10^6$)}  \\ \hline
\hline
392  & \Sn  &  3.2 $\pm$ 0.1 \\
662  & \Cs  &  22.5 $\pm$ 0.5 \\
1274 & \Na  &  22.3 $\pm$ 0.5 \\\hline
\end{tabular}
\caption{List of the radioactive sources imaged along with the gamma-ray energy and source strength. While \Na~provides two peaks, 511~keV and 1274~keV, only the imaging performance of the 1274~keV peak was assessed.}
\label{table:sourceStrength}
\end{table}

Due to the failure of three absorber bars, only four absorber bars, located centrally in the absorber plane, were included in the detector performance studies. 

 Long-range imaging tests were performed in a large room (16 m x 17 m) with a false floor at the National Research Council (NRC) in Ottawa, Canada. A list of the sources used is provided in Table~\ref{table:sourceStrength}. The gamma-ray emitting sources were placed 9~m away from the detector, at different angles [($\phi=0^{\circ}$,~$\theta=0,~10,~20,~30^{\circ}$),~($\phi=180^{\circ}$,~$\theta=10,~20^{\circ}$) and ($\phi=90^{\circ},~\theta=10,~20^{\circ}$)]. The origin of the coordinate system used lies at the center of the scatter plane. The x-axis lies along the length of the bars, the y-axis points upwards, and the z-axis points out of the front-face of the detector away from the absorber. The azimuthal angle, $\phi$, lies in the x-y plane and $\theta$ is the polar angle with respect to the z-axis.

\subsection{Event Selection}

\begin{figure}[!h]
\centering
\includegraphics[width=0.5\textwidth]{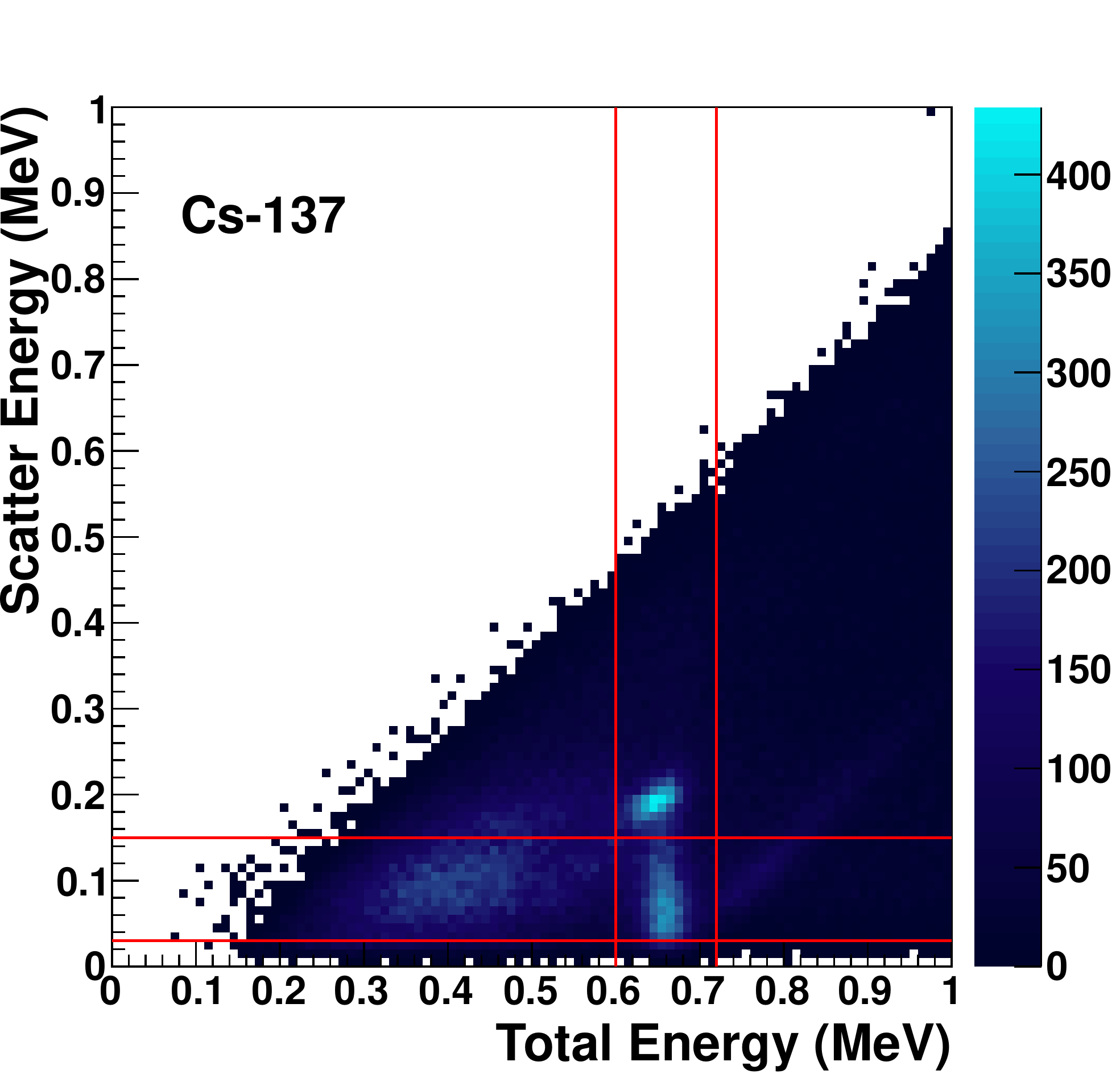}  
\caption{Scatter energy plotted against summed energy for \Cs~(662~keV). The vertical lines indicate the cuts performed on the summed energy. The horizontal lines indicate the selection performed on the scatter energy. The events that fall into the rectangular region formed by the overlap of the four lines were selected. Note that events from the back-scatter peak, which is visible at a total energy of 0.66~MeV and a scatter energy of 0.19~MeV, were thereby excluded.}
\label{fig:CsScatterPlotAll}
\end{figure}

A description of the trigger selection is provided in Section~\ref{sec:trigger}. Several criteria were used to select events off-line. First, events with exactly one scatter bar and exactly one absorber bar above threshold were selected. Second, a $\sim$ 2-$\sigma$ energy cut around the photopeak in the total energy distribution was implemented. Third, to reject back-scatter events, where a gamma ray is scattered in the absorber first, and then absorbed in the scatterer second, a $\sim$ 2-$\sigma$ exclusion cut was applied to the scatter-energy back-scatter peak. Fourth, a minimum threshold of 30~keV was required in the scatter detector. A summary of the energy selection used for the three sources is provided in Table~\ref{table:energycuts}. A visual representation of the cuts used for a \Cs~source is shown in Figure~\ref{fig:CsScatterPlotAll}.

\begin{table}[!h]                                                                                                                
\begin{center}                                                                                                               
\renewcommand{\arraystretch}{1.6}                                                                                            
\setlength{\tabcolsep}{7pt}                                                                                                 
\begin{tabular}{|c|c|c|c|c|c|}
\hline
{ Source} & {  Energy } & { Scatterer } & { Sum} \\
{} & {  (keV) } & { (keV) } & { (keV)} \\ \hline
\Sn & 392  &  30 - 100  & 340 - 440 \\
\Cs & 662 & 30 - 150  & 600 - 720 \\
\Na & 1272 & 30 - 180 or 280 - 650 & 1150 - 1350 \\
\hline
\end{tabular}
\end{center}
\caption{The event selection used includes a back-scatter rejection cut and a photopeak selection cut.}
\label{table:energycuts}
\end{table}

\begin{figure}[!h]
\centering
\includegraphics[width=0.5\textwidth]{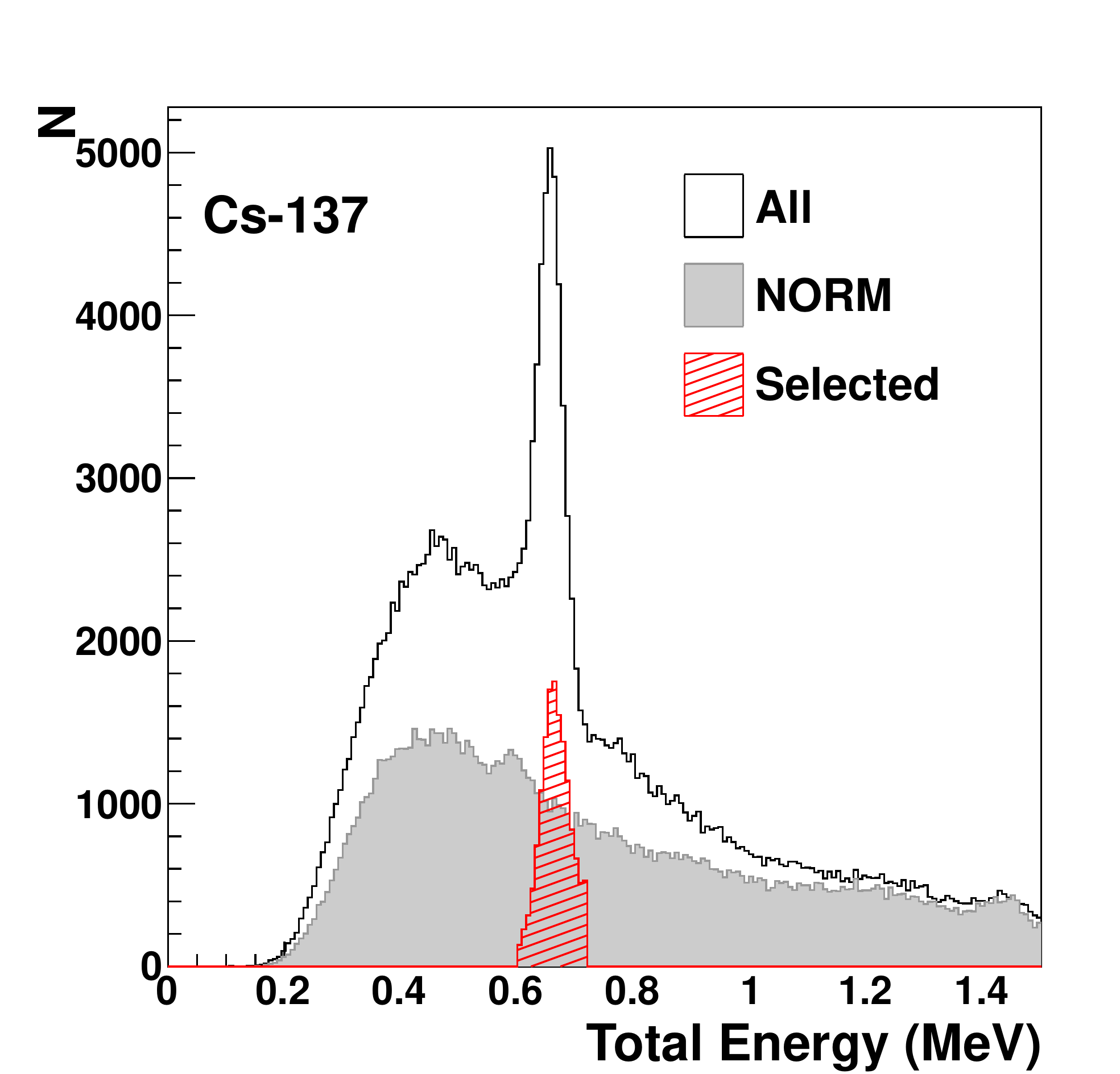}  
\caption{\Cs~(662~keV) energy distribution. White - all data. Grey Shaded - estimated contribution from NORM. Red Lined - selected events.}
\label{fig:CsEnergyDistTotal}
\end{figure}    

\begin{figure}[!h]
\centering
\begin{tabular}{cc}
\includegraphics[width=0.5\textwidth]{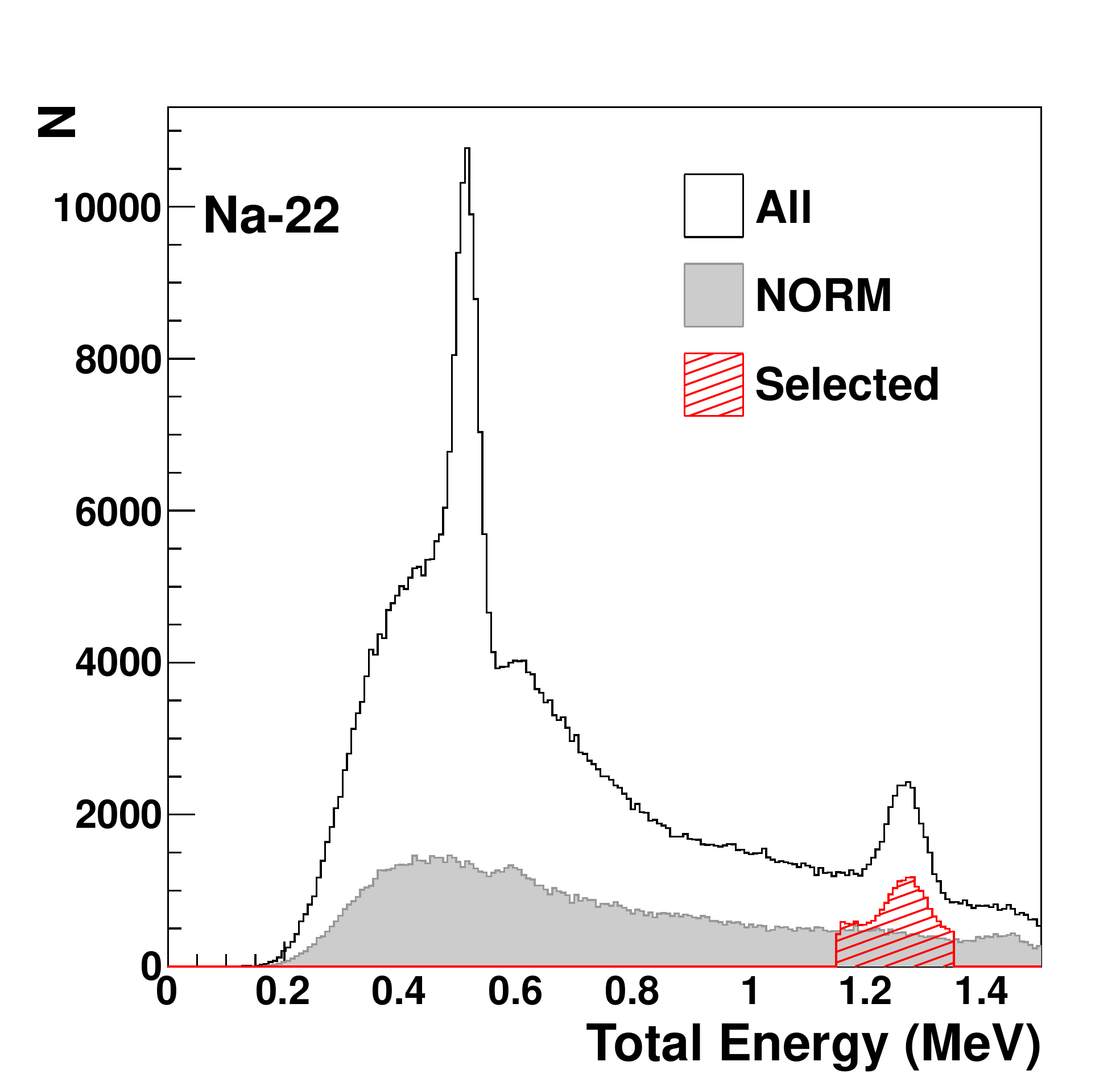}  
\includegraphics[width=0.5\textwidth]{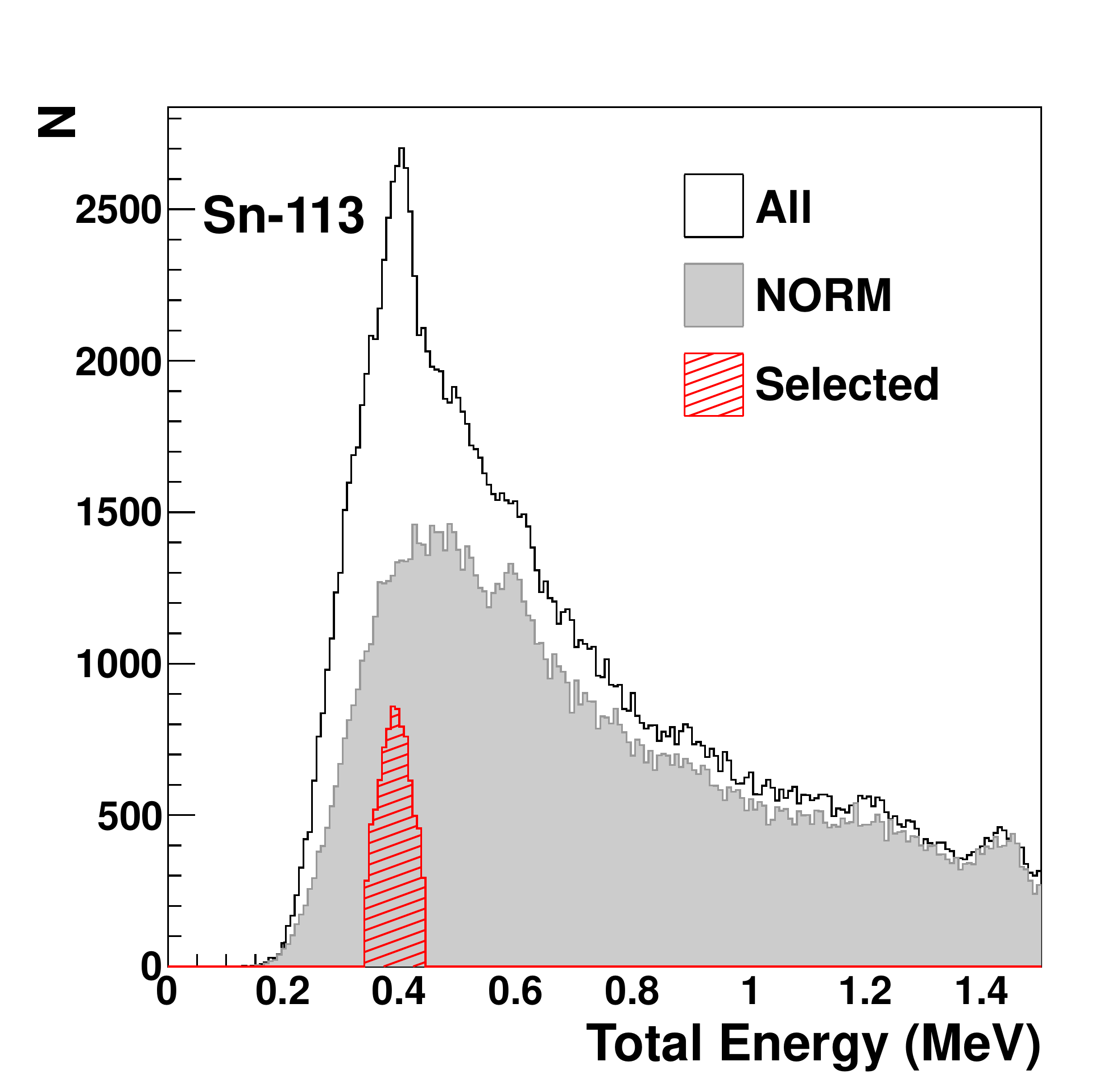}  
\end{tabular}
\caption{Energy distributions for \Na~(left) and \Sn~(right) from 2 hours of exposure. White - all data. Shaded Grey - estimated contribution from NORM.}
\label{fig:EnergyDist}
\end{figure}

 Figure~\ref{fig:CsEnergyDistTotal} shows the summed energy distributions (scatter + absorber) for a \Cs~source before and after the selection criteria are applied, overlaid with the estimated contribution from naturally occurring radioactive material (NORM). The NORM contribution was determined by taking a run of equal duration with no source present. Energy distributions for the \Na~and \Sn~sources are shown in Figure~\ref{fig:EnergyDist}. The varying fractional contribution from NORM to the total energy distribution is due to the different source strengths. There also exists a significant number of accidental coincidence events with NORM.

\begin{figure}[!h]
\centering
\includegraphics[width=0.5\textwidth]{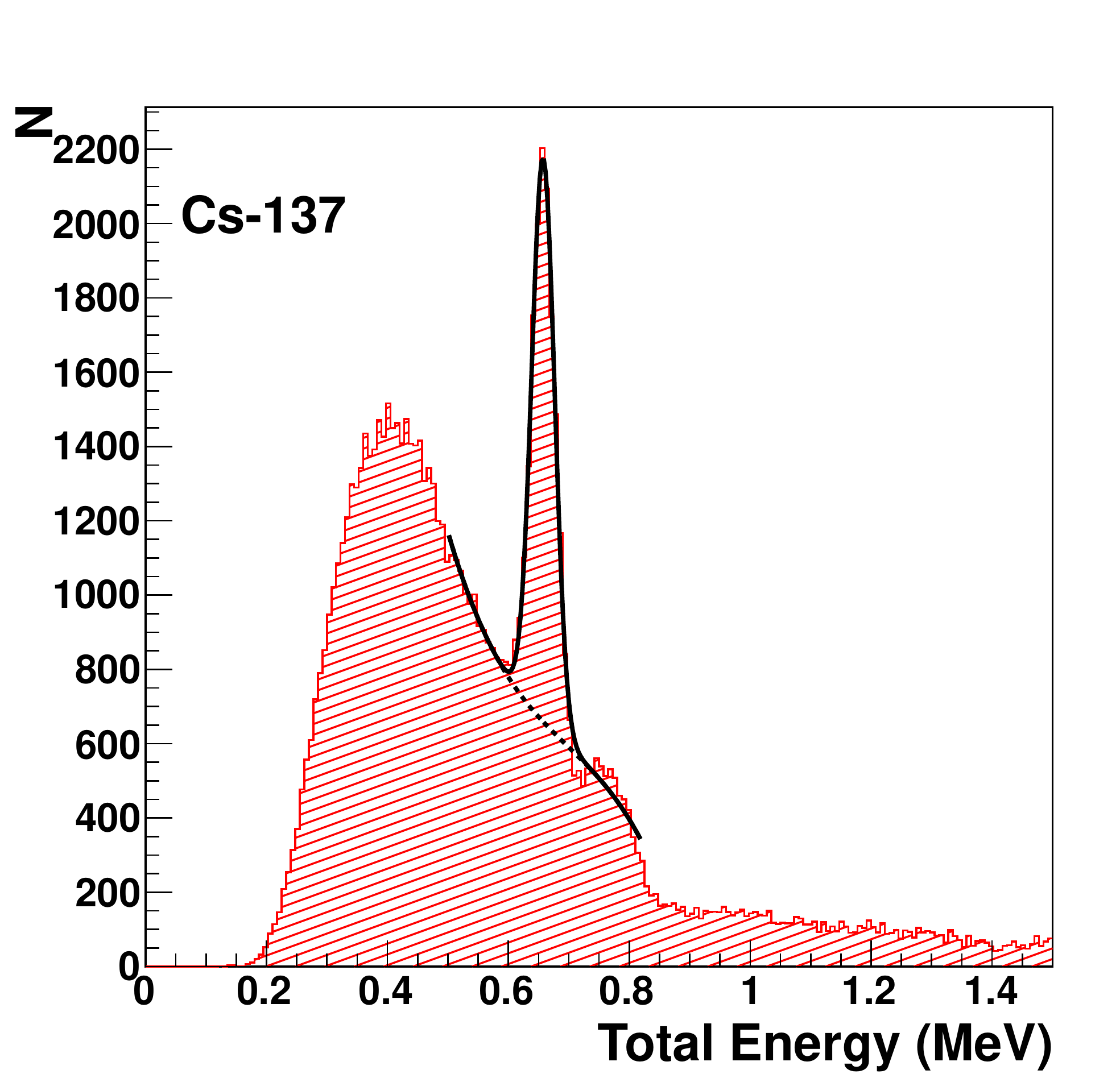}  
\caption{The summed energy distribution for \Cs~(662~keV) fit to a Gaussian and a third order polynomial.}
\label{fig:CsEnergyDistTotalFit}
\end{figure}

\begin{table}[!htp]
\centering
\renewcommand{\arraystretch}{1.3}
\setlength{\tabcolsep}{12pt}
\begin{tabular}{|c|c|}
\hline
{ Energy} & { Efficiency }\\
{ (keV)} & { ($\times 10^{-3}$)} \\ \hline
392 & 2.00 $\pm$ 0.09  \\
662 & 1.79 $\pm$ 0.04 \\
1274 & 1.10 $\pm$ 0.03 \\
\hline
\end{tabular}
\caption{Efficiency for three sources positioned at $\theta=10^{\circ}$, $\phi=0^{\circ}$. The uncertainties reflect both the statistical fluctuations and the uncertainty in the source emission rate.}
\label{table:efficiencies} 
\end{table}

We define the detector efficiency as ${N_{\mbox{\scriptsize}}}$/${N_{\mbox{\scriptsize Total}}}$, where $N_{\mbox{\scriptsize}}$ is the number of events that satisfy the event selection criteria and $N_{\mbox{\scriptsize Total}}$ is the total number of gamma rays passing through the active area of the front face of the detector (active area = 40 000 mm$^2$). To determine the efficiency, we relaxed the photopeak cut and in its place used a fit to isolate events in the photopeak. All of the other selection criteria were applied. To account for the number of background events, a Gaussian and a third order polynomial is fit to the summed energy distribution, see Figure~\ref{fig:CsEnergyDistTotalFit}. The correctly reconstructed forward-going scatters fall in the Gaussian part of the distribution. The efficiency of the detector for a 662~keV source placed off-axis ($\theta=10^{\circ}$) was measured to be $(1.79~\pm~0.04)~\times~10^{-3}$. The efficiencies at 392~keV and 1274~keV are provided in Table~\ref{table:efficiencies}.

\FloatBarrier

\subsubsection{Angular Resolution Measure}

\begin{figure}[!h]
\centering
\begin{tabular}{cc}
\includegraphics[width=0.45\textwidth]{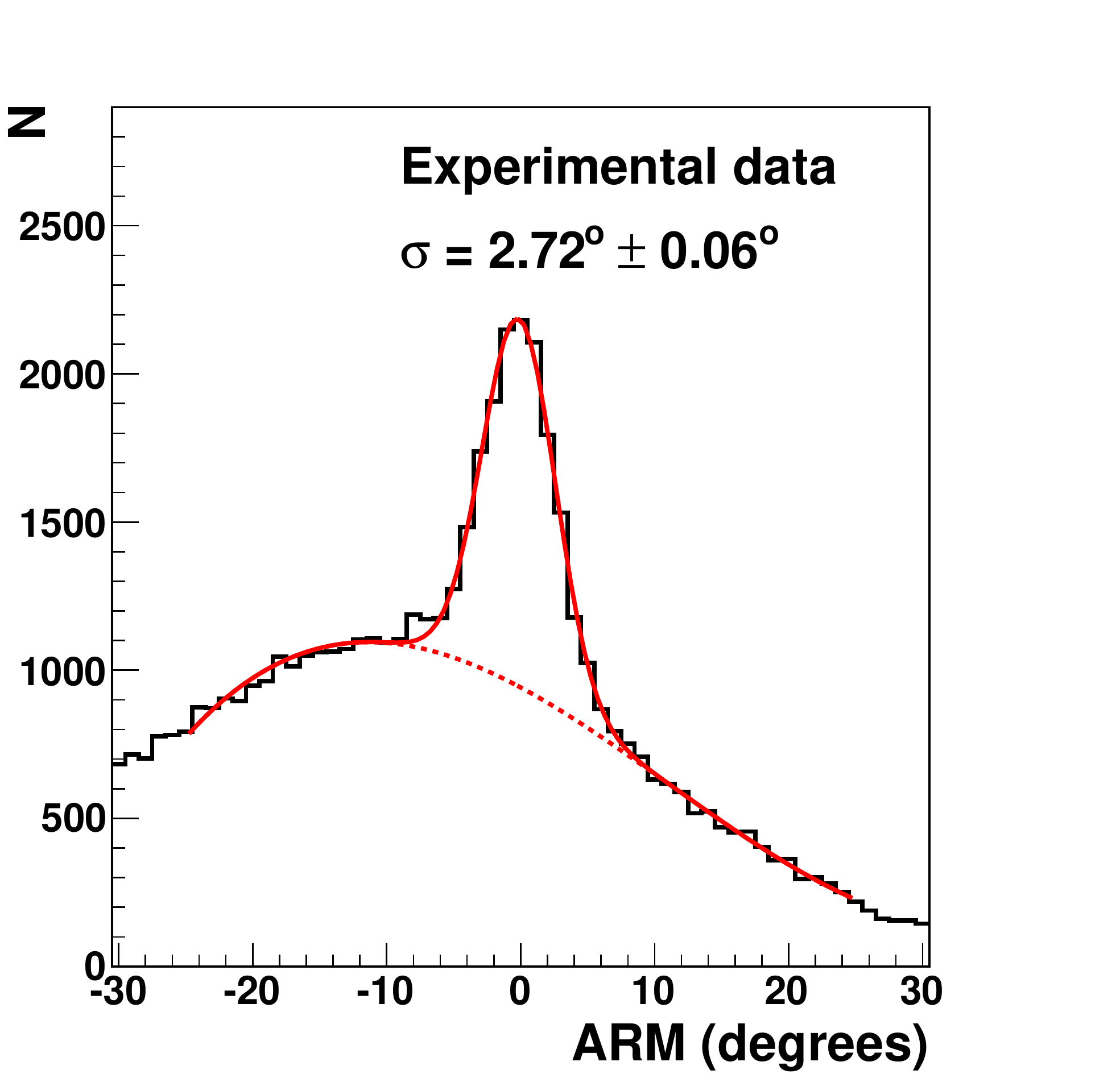}  &  
\includegraphics[width=0.45\textwidth]{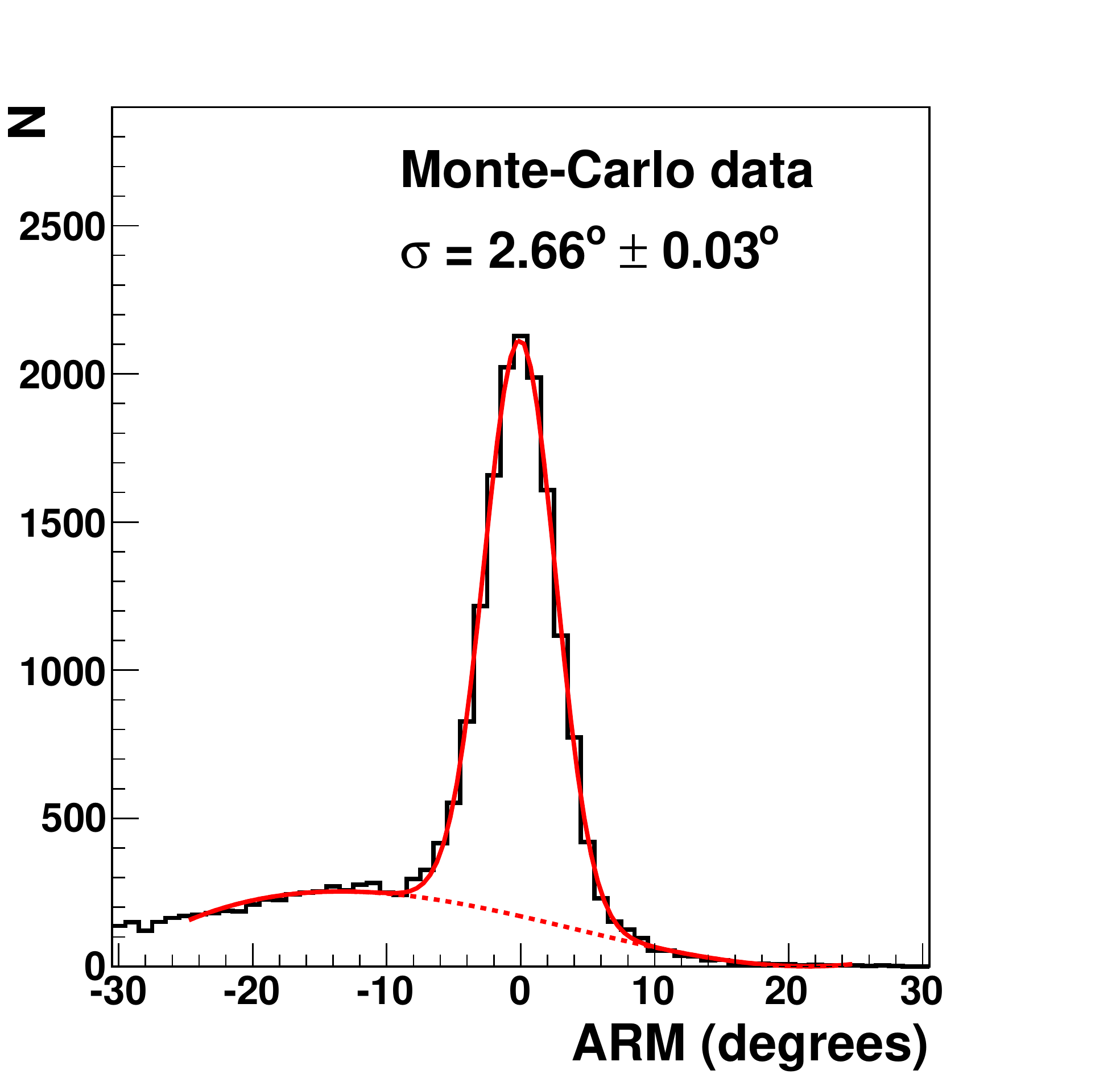}   
\end{tabular}
\caption{ARM distribution for a \Cs~source located on-axis at $\theta=0^{\circ}$ for selected events comparing measurement (left) and Monte Carlo (right). The distribution is fit to the sum of a Gaussian and a third-order polynomial. The widths of the Gaussian fits are shown.}
\label{fig:arm}
\end{figure}

 Figure~\ref{fig:arm} (left) shows the ARM distribution for the selected events of a \Cs~source located on-axis. The ARM distribution has been fit to the sum of a Gaussian distribution and a third-order polynomial. The width of the Gaussian fit yields \sigARM~of $2.72^{\circ}~\pm~0.06^{\circ}$.

The experimental ARM was compared with GEANT4 simulation. The detector configuration matched the imager that was tested, with ten scatter bars and four absorber bars. The simulation made use of realistic energy and position resolutions based on experimental data (see Section~\ref{sec: barperformance}). The recorded energies and longitudinal coordinates (x) of the interactions were smeared by Gaussian probability distributions with widths determined by the energy and position resolutions measured for each bar. Both the z and y-coordinates of the interactions were assumed to be at the bar centers. The simulated ARM distribution is shown in Figure 12 (right), and the width of its Gaussian fit is \sigARM~=~$2.66^{\circ}~\pm~0.03^{\circ}$. The good agreement with experimental data confirms our understanding of the detector performance.

\FloatBarrier
\subsection{Back-Projection Image}

\begin{figure}[!h]
\centering
\resizebox{0.45\textwidth}{!}{\includegraphics{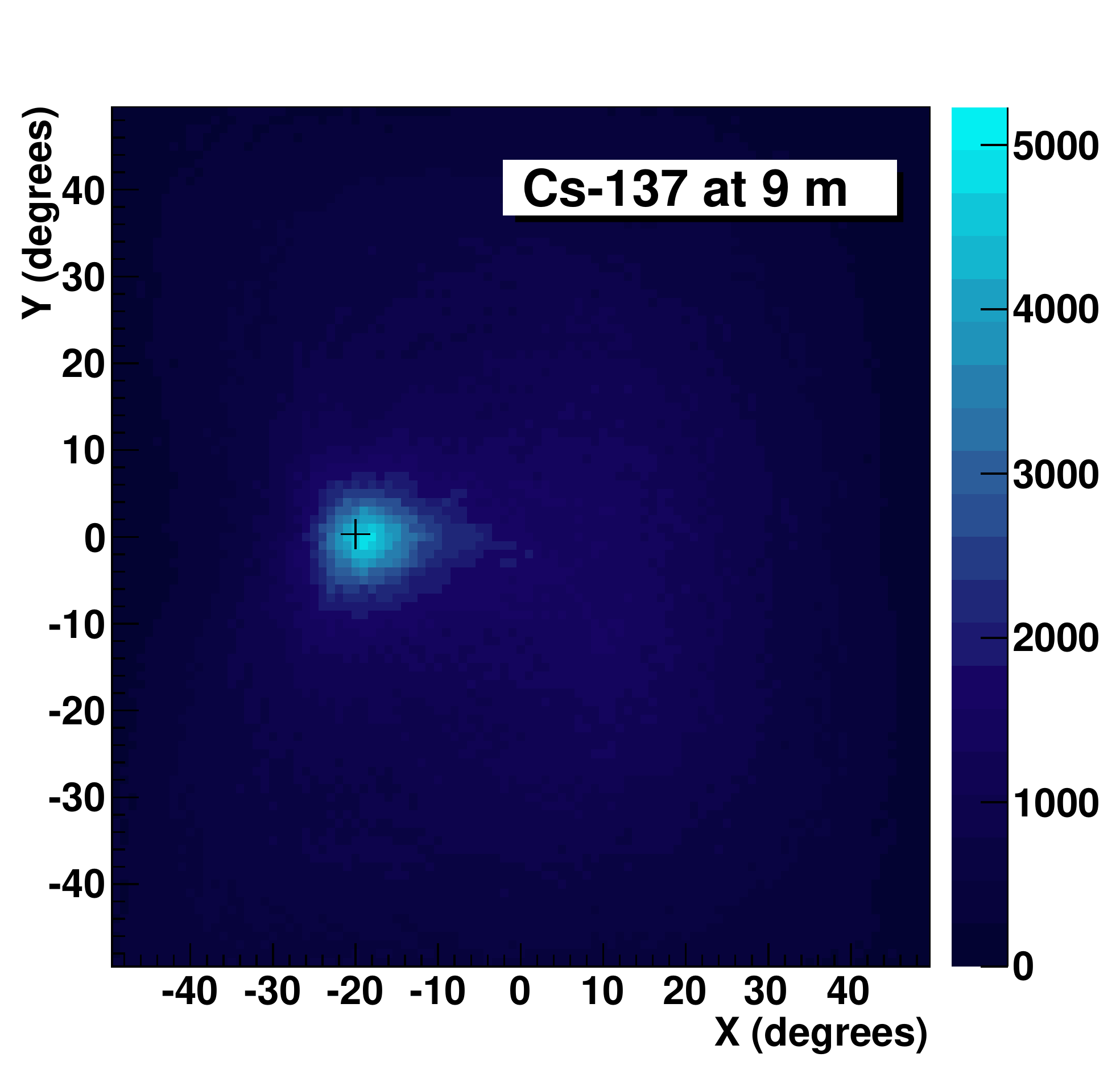}}
\caption{A back-projection image of selected events for a \Cs~source. The x and y-axis are presented in vertical and horizontal degrees. The true source position is indicated by cross-hairs. The broad excess at the origin is caused by NORM.}
\label{fig:backprojectionintro}
\end{figure}

  A back-projection image is a 2-D distribution formed by back-projecting Compton cones into angle space. Figure~\ref{fig:backprojectionintro} shows a back-projection image from $\sim$ 20~000 Compton cones (two hours of data) collected with the \Cs~source located 9~m away and 20$^{\circ}$ off axis. The back-projection image correctly indicates the position of the source.

\FloatBarrier
\subsection{$\chi^2$ Minimization Algorithm}

 An iterative $\chi^2$-minimization procedure using the MINUIT package~\cite{minuit} was employed to determine the vector $\hat{s}$($\theta$,$\phi$) which best represents the direction from the origin to the source. For $N$ events, a $\chi^2$ function is constructed with two fit parameters, $\theta$ and $\phi$, representing the polar and azimuthal angles of the source direction vector which are to be determined:

\begin{equation}
\chi^2=\sum_{i=1}^{N}\left[
  \frac{\mbox{ACA}(\theta,\phi)_i}{\sigma_{\mbox{\scriptsize ACA}i}}
  \right]^2~,
\label{eq_chi2}
\end{equation}
 where $\mbox{ACA}$($\theta$, $\phi$) is the angle of closest approach of the direction vector $\hat{s}$ to cone $i$, and $\sigma_{\mbox{\scriptsize ACA}i}$ is the uncertainty of this angle. The expression for $\mbox{ACA}$ is given by

\begin{equation}
\mbox{ACA} =  \arccos(\hat{s} (\theta,\phi) \cdot \hat{r_i}) - \theta^C_i~,
\label{aca}
\end{equation}
 where $\hat{r}_i$ is the unit-vector axis for cone $i$, and $\theta^C_i$ is the Compton scattering angle of the $i$th cone. The uncertainty of $\mbox{ACA}$ is calculated on an event-by-event basis from the cone-axis uncertainty and the Compton opening angle uncertainty (using Equation 1), taking into account the uncertainties on the energy deposits and their positions.

 To extract the source direction from a given sample of $N$ events, three iterations are performed, with the direction determined in each iteration passed on as a starting seed direction for the subsequent iteration. The highest bin of the coarse-binned back-projection image is used as a seed in the first iteration. For a detailed explanation of the algorithm see ref~\cite{SPIE}.

\clearpage
\FloatBarrier
\subsection{Imaging Performance at 662~keV}
\FloatBarrier

To characterize the imaging performance of the detector, the \Cs~source rates at a 9~m distance were converted into an equivalent 10~mCi source at 40~m with 100\% branching ratio\footnote[8]{The relative source strengths are taken into account, as is the $\frac{1}{r^2}$ dependence and the attenuation by the air.}. To quantify the ability to reconstruct $\theta$ and $\phi$, the events were divided into groups or trials of 60~s of data, and the $\chi^2$-minimization algorithm was applied to each group.  To determine the accuracy, the means of the resulting distributions of fitted $\theta$ and $\phi$ were used as a measure of the average reconstructed source location. To determine the angular precision, the 2-D standard deviation of the reconstructed direction was used. Table~\ref{tab:imagingperformance} summarizes $\avg{\theta}$, $\avg{\phi}$ and the angular precision obtained from sixty 60-second trials. The sources are all reconstructed to within about a degree of the known source position. The small discrepancy observed in the reconstructed angles versus the known source position can be explained by the fact that the quoted uncertainties on the reconstructed angles are statistical in nature and do not include a number of factors such as the effect of NORM, trigger threshold variations, and the depth of interaction in the bars. The angular precision obtained is on the order of a degree.

\begin{figure}[!htb]
\centering
\begin{tabular}{cc}
\includegraphics[width=0.45\textwidth]{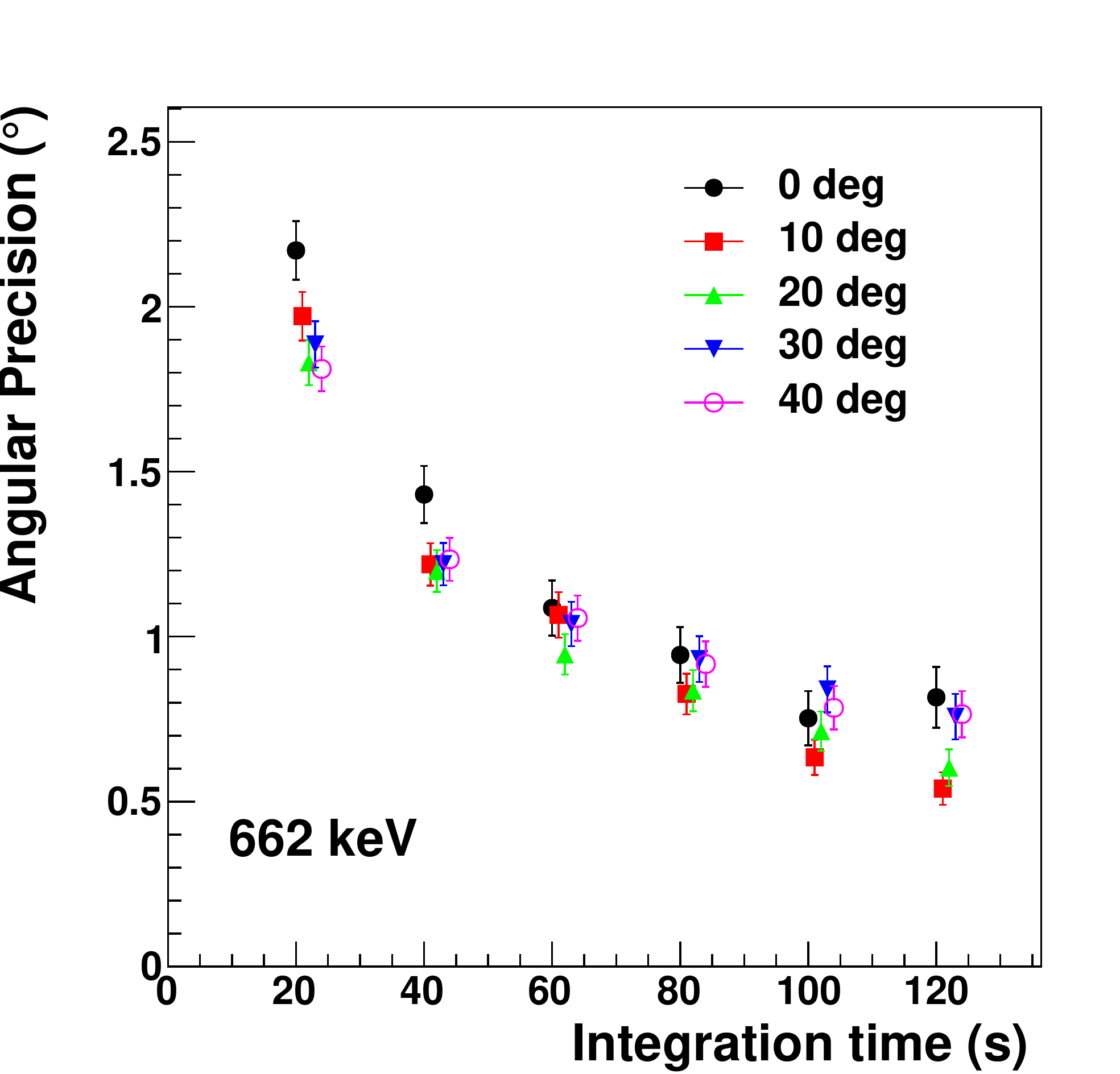}
\end{tabular}
\caption{Angular precision for an equivalent 10~mCi 662~keV source located 40~m away as a function of integration time for five source locations ($\phi$=0$^\circ$; $\theta$ = 0$^\circ$, 10$^\circ$, 20$^\circ$, 30$^\circ$, and 40$^\circ$). At all angles, the angular precision~reaches one degree in about a minute. Note that the data points are correlated and each point includes the entire data set.}
\label{fig:reconstrTTI}
\end{figure}

\begin{table}[!htb] 
\centering
\renewcommand{\arraystretch}{1.3}
\setlength{\tabcolsep}{6pt}
\begin{tabular}{|c|c|c||c|c|c|}
\hline
 { Run \#} & \multicolumn{2}{|c||}{ True Source Position} & \multicolumn{3}{|c|}{ Reconstructed Source Position}  \\ 
 & { $ \theta_{\tiny \mbox o} $} & {  $ \phi_{\tiny \mbox o} $ } & $\avg{\theta}$ & $\avg{\phi}$ & Angular Precision  \\
 & { ($^{\circ}$)} &  { ($^{\circ}$)}  & { ($^{\circ}$)}&  { ($^{\circ}$)} &  { ($^{\circ}$)} \\ \hline
 1 & 0.1  & 95.2  & 0.3 $\pm$ 0.2 & - & 1.1 $\pm$ 0.1  \\ \hline
 2 & 10.0 & 179.1 & 10.2 $\pm$ 0.1 & 180.0 $\pm$ 0.4 & 1.0 $\pm$ 0.1  \\ \hline
 3 & 20.0 & 179.6 & 21.4 $\pm$ 0.1 & 180.3 $\pm$ 0.2 & 1.1 $\pm$ 0.1  \\ \hline
 4 & 9.9  & 0.7   & 9.6 $\pm$ 0.2 & -0.9 $\pm$ 0.5 & 1.2 $\pm$ 0.1  \\ \hline
 5 & 20.0 & 0.3   & 19.7 $\pm$ 0.1 & -0.9 $\pm$ 0.2 & 1.0 $\pm$ 0.1  \\ \hline
 6 & 30.0 & 0.2   & 30.5 $\pm$ 0.1 & -0.8 $\pm$ 0.2 & 1.1 $\pm$ 0.1 \\ \hline
 7 & 40.0 & 0.2   & 39.6 $\pm$ 0.1 & -0.7 $\pm$ 0.2 & 1.1 $\pm$ 0.1  \\ \hline
 8 & 9.9  & 90.7  & 9.3 $\pm$ 0.2 & 91.1 $\pm$ 0.9 & 1.3 $\pm$ 0.1 \\ \hline
 9 & 20.1 & 89.3  & 20.6 $\pm$ 0.2 & 89.2 $\pm$ 0.4 & 1.4 $\pm$ 0.1 \\ \hline
\end{tabular}
\caption{Summary of the fit values obtained at 662~keV: $\avg{\theta}$, $\avg{\phi}$ and the angular precision for 60-second trials. Note that the source position for Run \#1 is very close to the z-axis, where $\phi$ is undefined, thus the $\phi$ value obtained for this run is not meaningful. }
\label{tab:imagingperformance}
\end{table}

\FloatBarrier

Figure~\ref{fig:reconstrTTI} shows the angular precision at 662~keV as a function of time for several source positions. 
The detector demonstrates a wide field-of-view with good performance for sources located up to 40~$^{\circ}$ off-axis.

\FloatBarrier
\subsection{Performance at Different Energies}

The 1274~keV and 392~keV peaks of a \Na~and \Sn~source located at $\theta=10^{\circ}$, $\phi=0^{\circ}$ were also successfully imaged. The results were again scaled such that the rates of gamma rays incident on the detector are equivalent to that from a 10~mCi source at 40~m with 100\% branching ratio, allowing for direct comparison with the 662~keV data. Due to the significant contribution from NORM to the \Sn~data, it was necessary to set the starting seed position used by the imaging algorithm to the known location of the source. For 60-second runs, we obtained an angular precision~of 1.0$^{\circ}$~$\pm$~0.1$^{\circ}$ and 1.6$^{\circ}~\pm~0.2^{\circ}$ for \Na~and \Sn,~respectively.

\FloatBarrier
\section{Conclusion}

We have developed a Compton-scatter gamma-ray imager composed of bars of NaI(Tl). The key to the low-cost design is the use of pulse-height sharing to reconstruct events, made possible by the successful development of a surface treatment that was applied to the bars. The technique was shown to be robust and the bars were incorporated into a fully-functional imager. The detector demonstrates good imaging performance over a wide range of energies (392~keV, 662~keV and 1274~keV) and over a wide field of view. An angular precision of about one degree has been achieved in one minute for a 10 mCi 662~keV gamma-ray source with 100\% branching ratio, located 40~m away.

We are currently investigating the use of different scintillator materials (CsI) and more compact readout devices, leveraging the experience our group has in constructing pixellated Compton imagers based on CsI(Tl) read out with silicon photomultipliers~\cite{Saullsipm2012,sinclairIEEE2013}.

\color{black}

\bibliographystyle{unsrt}
\bibliography{NIM_BarImagerMacLeod}







\end{document}